\documentclass[%
reprint, 
superscriptaddress,
 amsmath,amssymb,
  aps, 
 pre,
]{revtex4-1}
\usepackage{graphicx}
\usepackage{dcolumn}
\usepackage{bm}

\usepackage{color}

\newcommand{\pdiff}[2]{\frac{\partial #1}{\partial #2}}

\newcommand{\new}{\nonumber\\}
\newcommand{\abs}[1]{\left|#1\right|}

\newcommand{\erf}{{\rm erf}}
\newcommand{\hvar}{\hat{\varphi}}
\newcommand{\tvar}{\tilde{\varphi}}
\newcommand{\hA}{\hat{A}}

\newcommand{\hbeta}{\hat{\beta}}
\newcommand{\hphi}{\hat{\phi}}
\newcommand{\ox}{\underline{\bm{r}}}
\newcommand{\oy}{\underline{\bm{r}}^{\prime}}
\newcommand{\uxk}{\underline{\bm{r}}^{k}}
\newcommand{\uyk}{\underline{\bm{r}}^{\prime k}}
\newcommand{\dd}{\mbox{d}}

\newcommand{\br}{\bm{r}}
\newcommand{\bR}{\bm{R}}

\usepackage{amsmath}	

\RequirePackage[normalem]{ulem} 

\begin{document}

\preprint{APS/123-Qed} \title{Multiple glass transitions and higher
order replica symmetry breaking of binary mixtures}


\author{Harukuni Ikeda}
 \email{hikeda@g.ecc.u-tokyo.ac.jp}
\affiliation{
Graduate School of Arts and Sciences, The University of Tokyo 153-8902, Japan
}
\author{Kunimasa Miyazaki}%
\affiliation{%
 Department of Physics, Nagoya University, Nagoya, Japan
}%
\author{Hajime Yoshino}
\affiliation{%
 Cybermedia Center, Osaka University, Toyonaka, Osaka, Japan
}%
\affiliation{%
 Graduate School of Science, Osaka University, Toyonaka, Osaka, Japan
}%
\author{Atushi Ikeda}
 \email{atsushi.ikeda@phys.c.u-tokyo.ac.jp}
\affiliation{
Graduate School of Arts and Sciences, The University of Tokyo 153-8902, Japan
}%

\date{\today}
	     
\begin{abstract}
We extend the replica liquid theory in order to describe the multiple
glass transitions of binary mixtures with large size disparities, by
taking into account the two-step replica symmetry breaking (2RSB).  We
determine the glass phase diagram of the mixture of large and small
particles in the large-dimension limit where the mean-field theory
becomes exact. When the size ratio of particles is beyond a critical
value, the theory predicts three distinct glass phases; (i) the 1RSB
double glass where both components vitrify simultaneously, (ii) the 1RSB
single glass where only large particles are frozen while small particles
remain mobile, and (iii) a new glass phase called the 2RSB double glass
where both components vitrify simultaneously but with an energy
landscape topography distinct from the 1RSB double glass.
\end{abstract}


\maketitle 

\section{Introduction} 
Size dispersity of constituent atoms, molecules, or colloids is
ubiquitous in glassy systems. For most model glass formers employed in
numerical studies, the size dispersion is deliberately introduced in
order to avoid the crystallization. In experiments of colloidal or
polymeric glasses, it is simply difficult to eliminate. When the size
dispersity is small, it does not affect the nature of the glass
transition qualitatively; it only shifts the transition point or changes
the fragility slightly~\cite{Foffi2004,Tanaka2010}. However, if the size
dispersity is large, the nature of the glass transition qualitatively
and even dramatically changes. Due to the separation of the associated
length and time scales, dynamics of constituent particles with different
sizes decouple from each other~\cite{Angell2000}. A wide class of glassy
systems exhibit such decoupling phenomena, which include
ionic~\cite{martin1991}, metallic ~\cite{franz2003}, and polymeric
glasses~\cite{mayer2008,mayer2008m}, as well as colloidal
suspensions~\cite{imhof1995PRL,imhof1995,hendricks2015}. The simplest
model which shows the decoupling is the binary mixture of large and
small spherical particles with the disparate size ratio $R \equiv
\sigma_L/\sigma_S \gg 1$, where $\sigma_{L}$ and $\sigma_{S}$ are the
diameters of large and small particles, respectively. In the limit of
$R=\infty$, small particles behave as a solvent and only large particles
undergo the glass transition. As $R$ is reduced to the order of unity,
dynamics of small and large particles couple again and vitrify
simultaneously. The question is when and how the dynamics of the two
components decouple and the nature of the glass transition is altered as
$R$ is systematically changed~\footnote{We use a term ``decoupling'' to
mean the decoupling of the glass transition of small and large
particles, and not necessarily mean the decoupling of the self and
collective correlation functions. As shown later, the glass phase in the
infinite dimensional system is characterized solely by the self
correlation functions, not by the collective correlation functions.}.

Several experimental studies on binary colloidal
mixtures~\cite{imhof1995PRL,imhof1995,hendricks2015} have reported such
dynamical decoupling and the existence of multiple phases called the
\textit{single} glass where only large particles are frozen and
\textit{double} glass where both components vitrify simultaneously. But
the properties of different glass phases remain elusive. Several
simulation studies~\cite{moreno2006,moreno2006jcp,voigtmann2009} hint
the onset of the decoupling of the dynamics near the glass transition
point. However, the size ratios and time scales which can be covered by
simulations are limited. Currently, theoretical understanding of the
decoupling phenomena largely relies on the mode-coupling theory
(MCT)~\cite{bengtzelius1984,gotze2009}. Early studies have shown the
decoupling of dynamics of small and large particles
qualitatively~\cite{bosse1987,bosse1995} and a recent detailed analysis
predicted the emergence of rich multiple glass
phases~\cite{voigtmann2011}.  However, due to the series of uncontrolled
approximations inherent in the MCT, it is difficult to assess the
interplay of separate length scales and the validity of the theory.
Also, the other dynamical theory called self-consistent generalized
Langevin equation predicts a slightly different phase diagram for
binary mixtures in three 
dimensions~\cite{PhysRevE.99.042603,PhysRevE.100.042601}. One resolution
is to take the large-dimension limit where mean-field theories including
the MCT are expected to become exact, but the validity of the current
version of the MCT in this limit remains
controversial~\cite{Kirkpatrick1987b,Schmid2010,Ikeda2010,Jin2015,Maimbourg2016}.

In this work, we tackle this decoupling problem of the binary glasses
by constructing a statistical mechanical mean-field theory. Our theory
is based on the replica liquid theory
(RLT)~\cite{monasson1995,mezard1999,parisi2010}, which was originally
developed based on the classic mean-field spin-glass
theory~\cite{PhysRevB.36.8552,Kirkpatrick1987b,kirkpatrick1989,bouchaud2004}.
When the size dispersity is moderate, or the system is simply
mono-disperse, 
the output of the RLT can be summarized as follows. The
dynamic transition point which the MCT prescribes corresponds to the
\textit{spinodal} point in the RLT~\cite{Kirkpatrick1987b}. Beyond the
spinodal point, the RLT predicts the proliferation of exponentially
large number of metastable states, or minima, in the free energy
landscape. The logarithm of the number is the so-called configurational
entropy $\Sigma$. The RLT describes the {\it thermodynamic, or
ideal}, glass transition at the point where $\Sigma$
vanishes~\cite{kirkpatrick1989,monasson1995}. This transition is
accompanied by the one-step replica symmetry breaking (1RSB). This
scenario becomes exact in the mean-field (or large $d$)
limit~\cite{parisi2010,Kurchan2012}.

The RLT was extended to the binary mixtures, but it fails to predict the
decoupling phenomena even when the size ratio is large within the 1RSB
ansatz~\cite{coluzzi1999,coluzzi2000,biazzo2009,biazzo2010}. For the
spin-glass models which have the well-separated length/energy scales,
there arises the two-step replica symmetry breaking (2RSB) phase as the
stable solution~\cite{ikeda2016,crisanti2007,crisanti2011,crisanti2015},
which naturally captures the decoupling~\cite{ikeda2016}.  In this work,
we develop the RLT of the binary mixtures taking fully both one and two
step replica symmetry breakings into account. The new RLT predicts both
single and double glass phases and the physical mechanism can be
explained in the context of the energy landscape
picture~\cite{Goldstein1969,Sastry1998}. Interestingly, the new theory
also predicts a new glass phase which is characterized by the 2RSB
hierarchical structure of the free energy landscape.

\section{Replica method}
The replica method allows us to calculate the number of metastable
states and their entropic contribution to the free energy.  Before going
to the details of the model and calculations, here we briefly sketch the
main idea of the replica method.

\subsection{1RSB}
In the case of the 1RSB, we may simply label the free energy minima as
$\alpha=1,2,\ldots$. We introduce $m$ copies (replicas) of the original
system, which allows us to calculate the number of the minima, as
explained below. Then, the partition sum of the system is
~\cite{mezard1999,parisi2010}
\begin{align}
&Z_m=\sum_{\alpha}e^{-N m \beta f_{\alpha}}
 \new &=\int df e^{N (\Sigma(f)-\beta f)} = e^{N
(\Sigma(f^{*})-m \beta f^{*})},  \label{eq-ensemble-normal}
\end{align}
where $N$ is the number of particles, $\beta$ is the inverse temperature
$\beta=1/k_{\rm B}T$, and $f^{*}$ is the saddle point of the integration
over $f$, which can be calculated as
\begin{align}
 f^* =-\frac{1}{\beta}\pdiff{\log Z_m}{m}.
\end{align}
$\Sigma(f)$ is the configurational entropy,
\begin{align}
 \Sigma(f) \equiv \frac{1}{N} \log
\sum_{\alpha} \delta(f-f_{\alpha}).  
\end{align}
The saddle point value of $\Sigma$ can be calculated as follows~\cite{monasson1995}:
\begin{align}
&\frac{\log Z_m}{N}\approx -\beta m f^* +\Sigma(f^*)\new
&\to \Sigma(m) = \frac{\log Z_m}{N} + m \beta f^*
=-m^2 \pdiff{}{m}\left(\frac{\log Z_m}{mN}\right).\label{114745_24Oct20}
\end{align}
Finite configurational entropy $\lim_{m\to 1}\Sigma > 0$ means that the
system is glassy but still in the liquid state from the thermodynamic
point of view. Vanishing of the configurational entropy $\lim_{m\to
1}\Sigma \to 0$ means a thermodynamic transition to an ideal glass
state. Below the ideal glass transition point, $m=1$ is no longer the
most stable solution, and $m$ should be determined by $\Sigma(m)=0$,
leading to $m<1$. In the context of the 1RSB formalism, $m = 1$
corresponds to the liquid state, while $m< 1$ corresponds to the
thermodynamic glass
state~\cite{monasson1995,mezard1999,parisi2010}. The above scenario is
established theoretically in the limit $d \to
\infty$~\cite{parisi2010,Kurchan2012}.

\subsection{2RSB}
In the case of the 2RSB, it is natural to label different free energy
minimum with an index with two components
$\alpha=(\alpha_{L},\alpha_{S})$, where $\alpha_{L}=1,2,\ldots$
represents configurations of the large (L) particles. For each of such
configurations of the L particles, there are many configurations of the
small (S) particles for which we label as $\alpha_{S}=1,2,\ldots$.  This
hierarchical structure can be expressed by dividing $m$ replicas into
$m/m_1$ subgroups. The partition sum of the system can be expressed as
\begin{align}
&Z_m= \sum_{\alpha_{L}}\left(\sum_{\alpha_{S}\in \alpha_{L}}
 e^{-N m_1\beta f_{\alpha_{L},\alpha_{S}}}\right)^{\frac{m}{m_1}}
 =\sum_{\alpha_{L}} e^{-N
\frac{m}{m_1}\beta f_{\alpha_{L}}}, 
\new
& e^{-N\beta
f_{\alpha_{L}}} \equiv \sum_{\alpha_{S} \in \alpha_{L}} e^{-N
m_{1}\beta f_{\alpha_{L},\alpha_{S}}}.\label{152402_23Oct20}
\end{align}
In the last equation, we have introduced $f_{\alpha_{L}}$ which is the
free energy associated with a given configuration $\alpha_{L}$ of the L
particles obtained by taking a partial trace over the S particles. Here
the sum $\sum_{\alpha_{s} \in \alpha_{L}}$ denotes a trace over the
configurations of the S particles associated with a configuration
$\alpha_{L}$ of the L particles. The additional parameters $m$ and
$m_{1}$ are introduced as theoretical tools to detect the glass
transitions. Note that, in the special case $m=m_{1}$,
Eq.~(\ref{152402_23Oct20}) reduces back to the normal one
Eq.~(\ref{eq-ensemble-normal}). Then we are naturally lead to introduce two
kinds of configurational entropy associated with the L and S particles,
\begin{align}
&\Sigma_{1}(f) \equiv \frac{1}{N} \log \sum_{\alpha_{L}} \delta(f-f_{\alpha_{L}}),\new
&\Sigma_{2}(f;\alpha_{L}) \equiv \frac{1}{N} \log \sum_{\alpha_{s} \in
\alpha_{L}} \delta(f-f_{\alpha_{L},\alpha_{S}}).
\end{align}
The latter represents the number of energy minima with different
configurations of the S particles with a fixed configuration
$\alpha_{L}$ of the L particle. Repeating the similar argument of that
of the 1RSB, we can calculate the entropic contributions from
L particles $\Sigma_1$ and S
particles $\Sigma_2$ as~\cite{obuchi2010,ikeda2016}
\begin{align}
 &\Sigma_1 = 
 -m^2 \pdiff{}{m}\left(\frac{\log Z_m}{mN}\right),
 \new
 &\Sigma_2 = 
 -m_1^2 \pdiff{}{m_1\partial m}\left(\frac{\log Z_m}{mN}\right).\label{120152_24Oct20}
\end{align}
In the 2RSB case, thermodynamic glass transitions of the L and S
particles can take place either simultaneously or
separately. Physically, we anticipate the following two ideal glass
phases~\cite{obuchi2010,ikeda2016}. One possibility is that both the L
and S particles are in an ideal glass state such that
$\Sigma_{1}=\Sigma_{2}=0$ and thus $m<1$ and $m_{1}< 1$. The other
possibility is that only the L particles are in an ideal glass state
while the S particles still remain in the liquid state so that
$\Sigma_{1}=0$ but $\Sigma_{2}>0$ thus $m< 1$ but $m_{1}=1$.

\section{Model}
We consider a binary mixture of large (L) and small (S) spherical
particles interacting with a potential with a finite range, such as a
harmonic potential, given by $v_{\mu\nu}(r) =
\phi(r/\sigma_{\mu\nu})\theta(1-r/\sigma_{\mu\nu})$, where $\theta(x)$
is the Heaviside step function, $\mu,\nu\in \{L,S\}$ denotes the type of
particles, and $\sigma_{LL}$ and $\sigma_{SS}$ are the diameters of
large and small particles, respectively. We also assume that the
potential is additive, {\it i.e.},
$\sigma_{LS}=(\sigma_{LL}+\sigma_{SS})/2$. The reason to consider a
finite ranged potential is merely for technical simplicity; as shown
later, the functional form of $\phi(r)$ and temperature become
irrelevant parameters in the large-dimension limit.  There are only two
relevant parameters to characterize the thermodynamic phase diagram; the
volume fractions of each component $\varphi_\mu = N_\mu
V_d(\sigma_{\mu\mu})/V$ $(\mu \in \{L, S\})$, or equivalently the total
volume fraction $\varphi= \varphi_{S}+\varphi_{L}$ and the concentration
fraction (of small component), $x
=\varphi_{S}/(\varphi_{L}+\varphi_{S})$.  Here, $V_d(\sigma)$ is the
volume of a $d$-dimensional hypersphere with the diameter $\sigma$,
$N_\mu$ denotes the particle number of the $\mu$-component, and $V$ is a
volume of the system.  We represent the size ratio as
\begin{align}
\frac{\sigma_{LL}}{\sigma_{SS} } \equiv 1+ \frac{R}{d}, 
 \label{size-ratio}
\end{align}
so that the
volume ratio,
$V_d(\sigma_{LL})/V_d(\sigma_{SS})=(\sigma_{LL}/\sigma_{SS})^{d}$,
remains finite in the limit of $d\rightarrow \infty$.

\section{1RSB analysis}
\label{121359_24Oct20}

\begin{figure}[t]
\includegraphics[width=9cm]{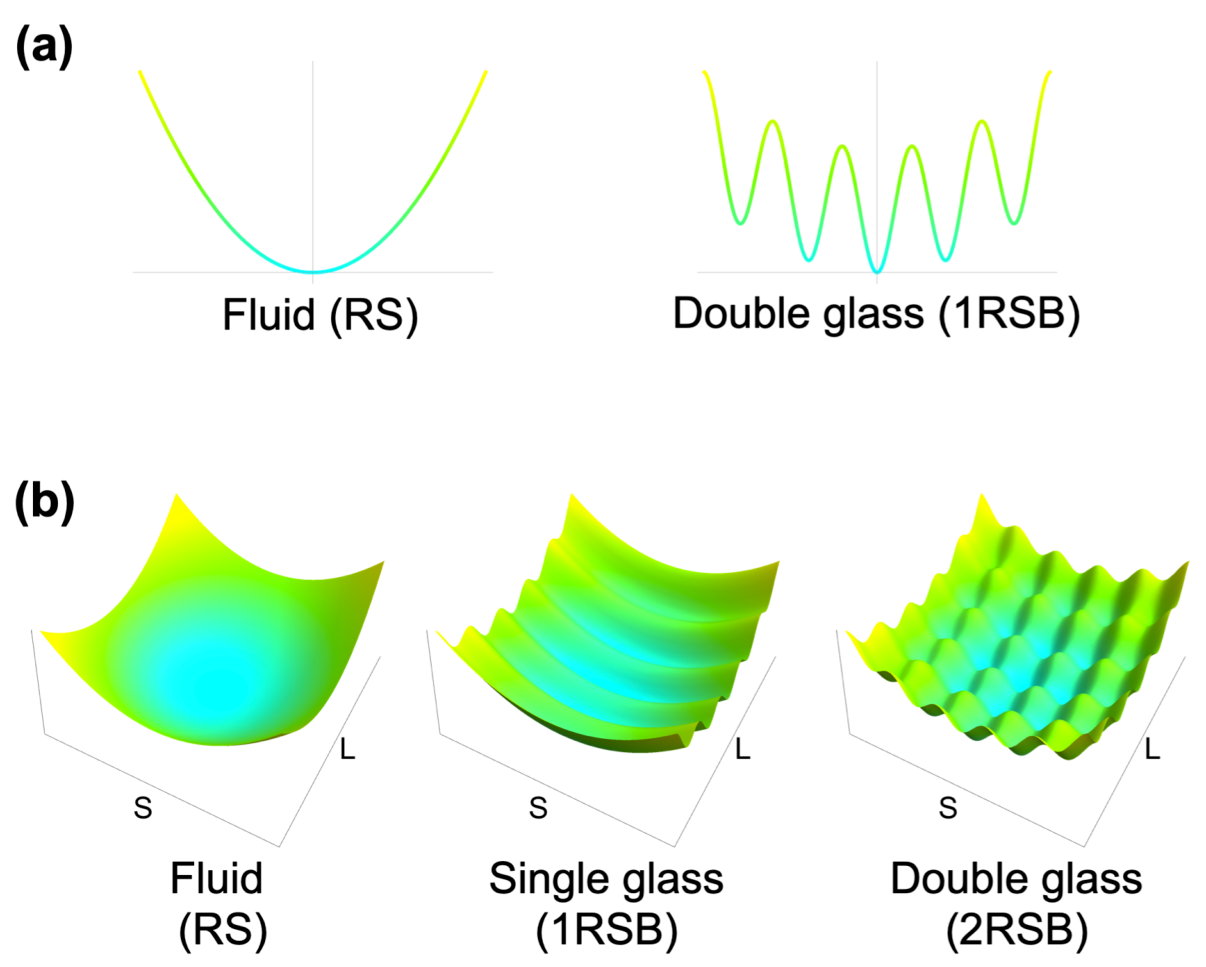} \caption{\small Schematic depiction of
the free energy landscape. (a) 1RSB formalism where the horizontal axis
represents the configuration of both large and small particles. (b) 2RSB
formalism where the coordinates labels ``L'' ans ``S'' represent the
configurations of large and small particles, respectively.  }
\label{fig_1}
\end{figure}
First, we consider the conventional RLT with the 1RSB ansatz (1RSB-RLT).
The landscape considered in the 1RSB formalism is schematically drawn in
FIG.~\ref{fig_1}~(a).  In the fluid, or the replica symmetric (RS)
phase, the free energy has a single minimum corresponding to equilibrium
fluid, while in the 1RSB phase, the free energy has multiple minima each
of which represents a different glass state. The main idea of the RLT is
to introduce the $m$ copies (replicas) of the original system to
distinguish the different glass states, which is referred to as the
overlap. The overlap, or similarity, between the configurations of
different replicas is finite within a same glass state, but zero between
different minima~\cite{monasson1995,mezard1999,biazzo2009,parisi2010}.
The 1RSB-RLT for monodisperse systems is well
developed~\cite{parisi2010} and the extension to binary mixtures is
straightforward~\cite{parisi2010,biazzo2009}, aside from a subtlety
related to the particle exchange in a glass
state~\cite{coluzzi1999,hikedanote,ozawa2016,ikeda2017,ikeda2019effect}. Following
the strategy of Ref.~\cite{parisi2010}, we introduce the density
distribution function in the replica space as density of
\textit{molecules} made of replicas,
\begin{align}
 &\rho_{\mu}(\ox) = \sum_{i\ {\rm for}\
 \mu\ {\rm particles}}\left\langle\prod_{a=1}^m\delta(\br^a-\br_i^a)\right\rangle,
 \label{170442_8Oct20} 
\end{align} 
where $\ox=\{\br^1,\cdots,\br^m\}$ represents the set of the particle
positions in the replica space~\cite{biazzo2009,parisi2010}, and $\mu\in
\{L,S\}$ denotes the particle species. Expanding the free energy of the
replicated system by $\rho_\mu(\ox)$, we obtain
\begin{align}
 \log Z_m &= 
 \sum_{\mu\in\{L,S\}}\int  d\ox \rho_\mu(\ox)(1-\log\rho_\mu(\ox)),\new
 &+\sum_{\mu\nu\in \{L,S\}}\frac{1}{2}\int d\ox d\oy\rho_\mu(\ox)\rho_\nu(\oy)f_{\mu\nu}(\ox-\oy)+O(\rho^3),\label{160828_18Mar16}
\end{align}
where we intoduced the Mayer function defined by
\begin{align}
 f_{\mu\nu}(\ox-\oy) &= \prod_{a=1}^m e^{-\beta v_{\mu\nu}(\br^a-{\br'}^a)} -1.\label{164257_16Sep17}
\end{align}
In the $d\to\infty$ limit, the $O(\rho^3)$ term is negligible
compared to the first and second order terms~\cite{parisi2020theory}.
Furthermore, in this limit, only the first and second cumulants of
$\rho_\mu(\ox)$ are relevant~\cite{parisi2020theory}, meaning that
Eq.~(\ref{170442_8Oct20}) can be represented by the Gaussian function
as~\cite{biazzo2009,parisi2010}
\begin{align}
 \rho_{\mu}(\ox) &= \rho_\mu\int d\bR \prod_{a=1}^m\gamma_{A_\mu}(\br^a-\bR),\label{161925_18Mar16}
\end{align}
where $\gamma_A(\bm{r})=(2\pi A)^{-d/2}e^{-\frac{\abs{\br}^2}{2A}}$.
$A_\mu$ represents the strength of the correlation between $m$
replicas. This is to be determined by the saddle point condition, as
described below. Substituting Eq.~(\ref{161925_18Mar16}) into
Eq.~(\ref{160828_18Mar16}), we obtain
\begin{align}
&\log Z_m
 = \sum_{\mu}N_\mu\left[-\frac{d}{2}(1-m)\log 2\pi A_\mu
 -\frac{d}{2}(1-m-\log m)\right]\new
&+ \sum_{\mu}N_\mu \left(1-\log\rho_\mu \right)
 + \frac{1}{2}\sum_{\mu\nu}\frac{N_\mu N_\nu}{V}\int d\br \left(q_{\mu\nu}^m(\br)-1\right),\label{212601_18Mar16}
\end{align}
with
\begin{align}
 q_{\mu\nu}(\br) &= \int d\bR\gamma_{A_{\mu}+A_\nu}(\br+\bR)e^{-\beta v_{\mu\nu}(\bR)}.
\end{align}

\begin{figure}[t]
\includegraphics[width=9cm]{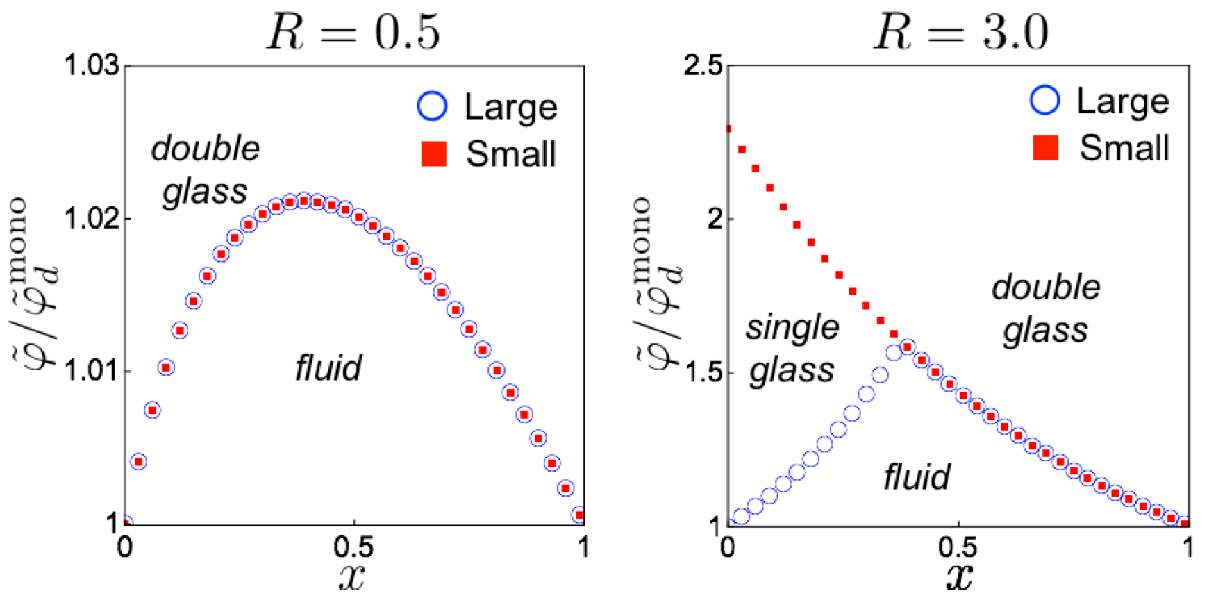}
\caption{\small Dynamic phase diagram for $R=0.5$ and $R=3.0$: The
circles and filled squares denote the dynamical transition points of
large and small particles, respectively.  The $y$-axis is scaled by the
dynamical transition point of the monodisperse system,
$\tilde{\varphi}_d^{\rm mono}$.} \label{230233_18Mar16}
\end{figure}

We first calculate the dynamical transition point $\varphi_d$ of our
model at which the non-trivial solution of $A_\mu$ arises.  $A_\mu$ can
be calculated from the saddle point condition $\partial_{A_\mu}\log
Z_m=0$ by taking $m\to 1$ limit for fixed $d$ and then taking the limit
of $d\to\infty$ afterwards.  For simplicity, below, we investigate the
hard-sphere limit ($T\rightarrow 0+$). The result of monodisperse hard
spheres in the large-dimension limit~\cite{parisi2010} can be easily
extended to binary mixtures. The saddle point condition leads to
\begin{align}
 \frac{1}{\hA_L} &\sim  \tvar \left[(1-x)M(\hA_L)+ x e^{r/2}M\left(\frac{\hA_L+\hA_S}{2}\right)\right],\new
 \frac{1}{\hA_S} &\sim \tvar \left[x M(\hA_S)+ (1-x)e^{-r/2}M\left(\frac{\hA_L+\hA_S}{2}\right)\right],
 \label{225822_18Mar16}
\end{align}
where we defined $\hA_\mu=d^2 A_\mu/\sigma_{\mu\mu}^2$ and introduced
the reduced density as
\begin{align}
\tvar=\frac{2^d\varphi}{d}.
\end{align}
$M(\hA)$ is the auxiliary function defined by
\begin{align}
 M(\hA) &= -\int dy e^y \log \left[\Theta\left(\frac{y+\hA}{\sqrt{4\hA}}\right)\right]
 \pdiff{}{\hA}\Theta\left(\frac{y+\hA}{\sqrt{4\hA}}\right),\new
 \Theta(x) &= \frac{1}{2}\left(1 + \erf(x)\right).
\end{align}
Solving Eqs.~(\ref{225822_18Mar16}) numerically, we obtain $A_\mu$ and
$\varphi_d$. We show the resultant (dynamic) phase diagram in
FIG.~\ref{230233_18Mar16}.  When the size ratio between large and small
particles, $R$ defined by Eq.~(\ref{size-ratio}), is small, the
dynamical transitions of large and small particles take place
simultaneously (see the left panel of FIG.~\ref{230233_18Mar16}). One
observes only the double glass phase in which all particles are frozen.
Contrarily, if $R$ is sufficiently large ($\geq R_c\approx 0.6$), the
glass phase splits into the two phases. See the right panel of
FIG.~\ref{230233_18Mar16}.  At small $x$ and at modelete densities, one
obtains the single glass phase in which only large particles are
frozen. As the density further increases, small particles undergo the
dynamical transition and enter to the double glass phase.  At large
$x$'s, on the contrary, the system enters to the double glass phase from
the fluid phase without bypassing the slngle glass phase.

\begin{figure}[t]
 \includegraphics[width=9cm]{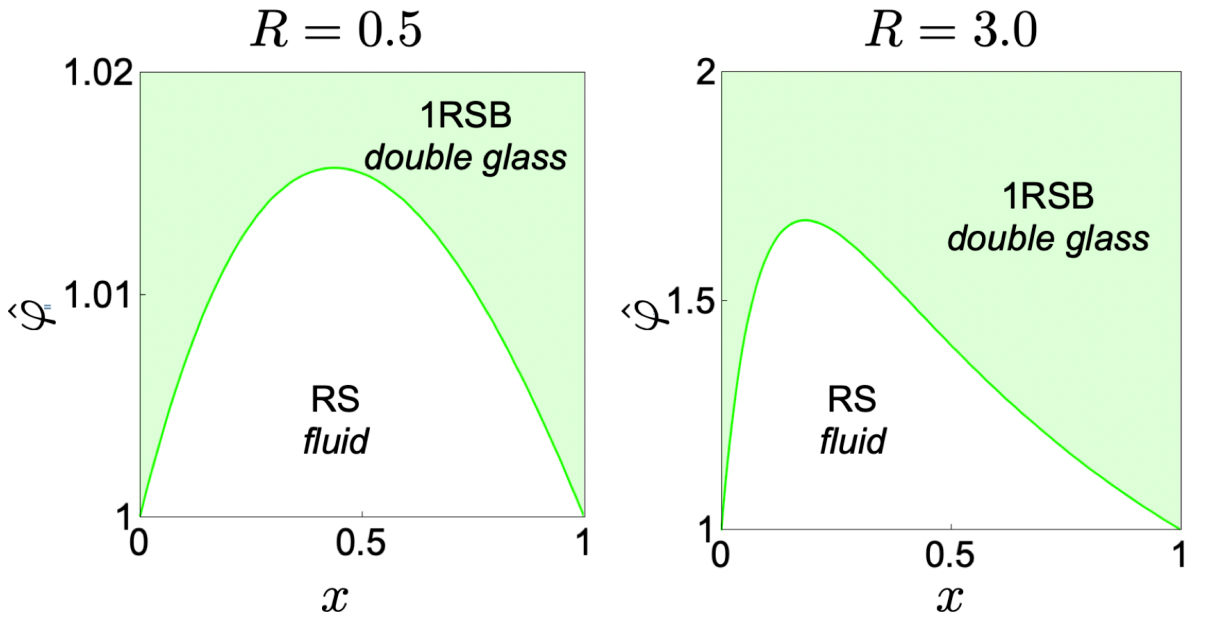}
\caption{\small 1RSB phase diagram for $R=0.5$ and $R=3$: The solid line
denotes the normalized thermodynamic glass transition point,
$\hat{\varphi}_K$.}  \label{fig_2}
\end{figure}

Next, we discuss the thermodynamic glass transition point, $\varphi_K$,
where the configurational entropy vanishes. In the large-dimension
limit, the thermodynamic glass transition density scales as
$\varphi_K=O(2^{-d}d\log d)$ and the cage size scales as
$A_\mu=O(1/d^2\log d)$~\cite{parisi2010}, leading to
$\gamma_{A_{\mu}}(x)\sim \delta(x)$ and $q_{\mu\nu}(r)\sim e^{-\beta
v_{\mu\nu}(r)}$. Substituting this into Eq.~(\ref{212601_18Mar16}), we
obtain the asymptotic form of the free energy near the thermodynamic
glass transition point,
\begin{align}
 \frac{\log Z_m}{Nm} &= \frac{1}{m}\left[\frac{d}{2}\log d
 -\frac{\varphi}{2}\frac{(1-x+xe^{R/2})^2}{1-x+xe^R}I(m)
 \right]-d\log d,
\end{align}
where
\begin{align}
I(m) &=\int_{-\infty}^{\infty}\!\!\dd y~ e^y\left[1-e^{-m\hbeta\hphi(y)}\right]
\end{align}
with $\hphi(y) = d^2\phi\left(1+{y}/{d}\right)$.  
$\varphi_K$ is calculated by $\lim_{m\to
1}\Sigma(m)=0$~\cite{parisi2010}, where the configurational entropy
$\Sigma(m)$ is give by Eq.~(\ref{114745_24Oct20}). After some
manipulations, we obtain
\begin{align}
\hvar_K(x) &=\frac{\varphi_{K}(x)}{\varphi_K^{\rm mono}} =
 \frac{1-x+xe^R}{(1-x+xe^{R/2})^2}\label{1rsb},
\end{align}
where $\varphi^{\rm mono}_K=2^{-d}d\log d/h(1)$ 
with
$h(m)=-m^2\partial_m (I(m)/m)$ 
denotes the thermodynamic glass transition density for
the one-component system. Eq.~(\ref{1rsb})
implies that the thermodynamics transition point 
does not depend on
$\beta$ and $v_{\mu\nu}$,
if one uses the reduced density
\begin{align}
 \hat{\varphi} = \frac{\varphi}{\varphi_K^{\rm mono}}.
\end{align}
Typical phase diagrams predicted by Eq.~(\ref{1rsb}) are shown in
FIG.~\ref{fig_2}. As mentioned before, the 1RSB RLT fails to describe
the decoupling of the thermodynamic glass transition points of large and
small particles.

\section{2RSB analysis}

Next, we introduce the 2RSB ansatz into the RLT, motivated by the recent
study of a binary version of mean-field spin-glass
model~\cite{ikeda2016}. To discuss the decoupling, we separately
consider the configuration of large and small particles, as shown in
FIG.~\ref{fig_1}~(b). For sufficiently small $\varphi$, the system is in
the RS fluid phase, where the free energy has a single minimum
(FIG.~\ref{fig_1}~(b) left). For intermediate $\varphi$'s, there appears
the single glass phase described by the 1RSB, where large particles are
vitrified while small particles are mobile. The minimum of the free
energy splits into multiple glass states due to the configuration of
large particles (FIG.~\ref{fig_1} (b) middle). For sufficiently large
$\varphi$, small particles also vitrify, as a consequence, the minima
further split due to the emergence of the multi-valley for the
configuration of small particles (FIG.~\ref{fig_1} (b) right). Only the
2RSB formalism can describe this hierarchical structure and the
decoupling of the thermodynamic glass transition points of large and
small particles. The RLT with the 2RSB ansatz is formulated by dividing
$m$ replicas into $m/m_1$ sub-groups, each of which contains $m_1$
replicas. The $m_1$ replicas of small particles within a same sub-group
are constrained around their center of mass, whereas the replicas of
different sub-groups can move independently. For large particles, all
$m$ replicas are constrained around their center of mass. In other
words, the replicated liquid is a $(m/m_1+1)$-component
\textit{molecular} mixture which consists of $m/m_1$ types of molecules
composed of $m_1$ small particles and one type of molecules composed of
$m$ large particles. Note that the higher order RSB is a natural
consequence of consecutive transitions of each component, and this
picture is distinct from the full RSB transition recently studied in the
context of the \textit{marginal} glass transition where each RSB state
corresponds to one frozen state~\cite{charbonneau2014}.

Based on the 2RSB ansatz, one can write down the free energy of the
replica liquid using the virial expansion of the standard grand
canonical partition function, which reads
\begin{widetext}
\begin{align}
&\log Z_m =
\int\dd\ox \rho_L(\ox)(1-\log\rho_L(\ox)) +\sum_{k=1}^{m/m_1}
 \int\dd\uxk \rho_{S_k}(\uxk)(1-\log \rho_{S_k}(\uxk)) 
+ \frac{1}{2}\int\dd\ox d\oy\rho_L(\ox)\rho_L(\oy)f_{LL}(\ox-\oy) \new
& + \sum_{k=1}^{m/m_1} \frac{1}{2}\int\dd\uxk d\uyk\rho_{S_k}(\uxk)\rho_{S_k}(\uyk)f_{SS}(\uxk-\uyk)
+\sum_{k=1}^{m/m_1} \int\dd\ox d\uyk\rho_L(\ox)\rho_{S_k}(\uyk)f_{LS}(\uxk-\uyk) 
+ O(\rho_L^3,\rho_{S_k}^3).
\label{160118_22Feb16} 
\end{align}
\end{widetext}
In this expression, $\rho_{S_k}$ is the density field of small particles
of the $k$-th type ~\footnote{Note that $\rho_\alpha$ have only the
information of the \textit{tagged} variables and not of the
\textit{collective} variables.}. $\ox=\{\br^1,\cdots,\br^m\}$ and
$\uxk=\{\br^{1,k}, \cdots, \br^{m_1, k}\}$ represent their coordinates in
the replica space.  $f_{\mu\nu}(\ox-\oy)$ $(\mu,\nu\in L, S_k)$ is the
Mayer function defined by
\begin{align}
 f_{\mu\nu}(\ox-\oy) &= \prod_{a} e^{-\beta v_{\mu\nu}(\br^a-\br^{\prime,a})} -1,
\end{align}
where the product over $a$ is made only for the replicas commonly
included in the $\mu$ and $\nu$ molecules. The first and second terms of
Eq.~(\ref{160118_22Feb16}) are the ideal gas parts and the third to
fifth terms represent the interaction contributions~\cite{parisi2010}.
As before, we give the profiles of $\rho_L(\ox)$ and $\rho_{S_k}(\uxk)$
as Gaussian~\cite{parisi2010}, which is known to be exact in the
large-dimension limit~\cite{Kurchan2012}.

It should be emphasized that, in the large-dimension limit, only the lowest
order term in the Mayer expansions survives, which simplifies the
analysis considerably. This implies that, in the
large-dimension limit, the
so-called \textit{depletion force}, a short-ranged attraction between
large particles induced by small ones, is
absent~\cite{asakura1954,dijkstra1999}, which is intrinsically the
higher order effect.

The glass phases are determined by optimizing the free energy,
Eq.~(\ref{160118_22Feb16}), with respect to $m$,  $m_1$, and the cage
sizes. The thermodynamic
glass transition density $\varphi_K $ is the point at which
the configurational entropy vanishes. In the vicinity of $\varphi_K\sim O(d\log
d)$~\cite{parisi2010}, the free energy can be simplified and written by
an asymptotic expression
\begin{align}
\frac{\log Z_m}{Nm} &=  g_1(m) + g_2(m_1)-d\log d, 
\label{153441_11Oct15}
\end{align}
with the auxiliary functions defined by
 \begin{align}
g_1(m) &= \frac{1}{m}\bigg{[}
\frac{1-x}{1-x+x e^R}\frac{d}{2}\log d
 -\frac{2^d\varphi}{2}\frac{(1-x)^2}{1-x+x e^R}I(m)
\bigg{]},\new
g_2(m_1) &= \frac{1}{m_1}\bigg{[}
\frac{x e^R}{1-x+x e^R}\frac{d}{2}\log d \new
&-\frac{2^d\varphi}{2}\frac{x^2e^R+2x(1-x)e^{R/2}}{1-x+x e^R}I(m_1)
\bigg{]}.\label{212753_18Mar16}
\end{align}
Inside the glass phases, $\varphi > \varphi_{K}$, $m$ and $m_1$ become
smaller than unity. There are two possibilities, $m<m_1<1$ and
$m=m_1<1$, which should be treated separately.

\begin{figure}[bt]
\includegraphics[width=1.0\columnwidth]{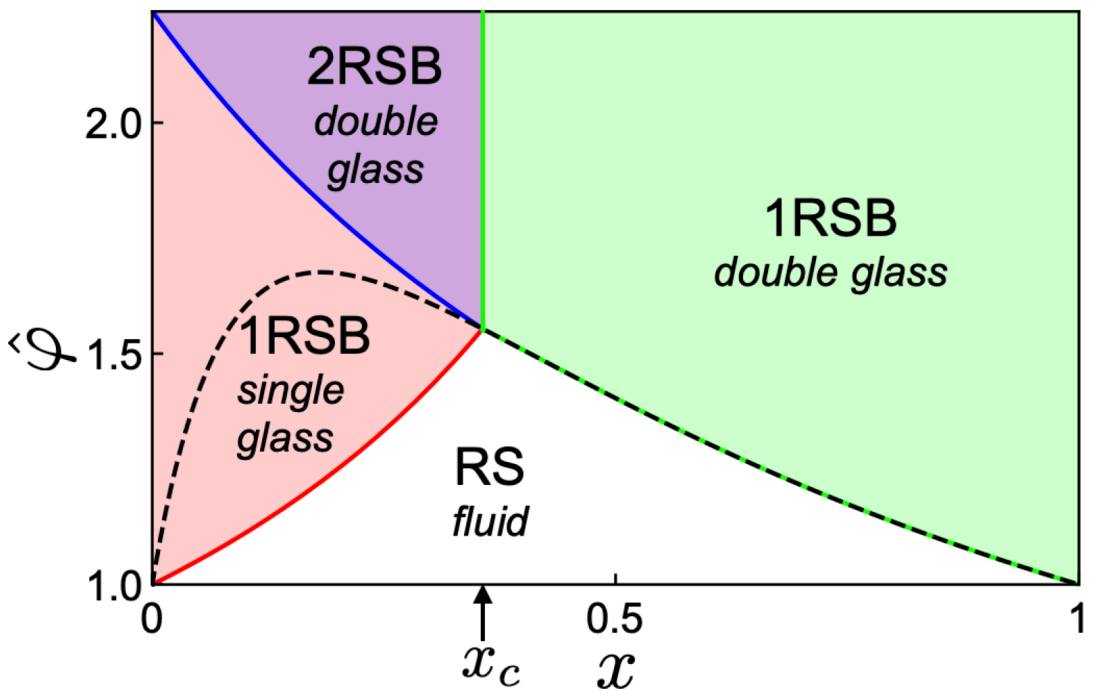} \caption{\small{Full
 phase diagram for $R=3$. $x$ is the concentration fraction of small
 particles. $\hvar$ is the packing fraction divided by the glass
 transition point of the one-component system.  The broken line is the
 glass transition line obtained by the 1RSB-RLT.
}} \label{fig_4}
\end{figure}
In the case of $m<m_1<1$, the glass phase is characterized by the 2RSB
free energy (FIG.~\ref{fig_1} (b)). $m$ and $m_1$ are determined by
solving the saddle point equations, $\Sigma_1=0$ and
$\Sigma_2=0$. Substituting Eq.~(\ref{153441_11Oct15}) into
Eqs.~(\ref{120152_24Oct20}), we get
 \begin{align}
  \frac{h(m)}{h(1)} &= \frac{\varphi_K^{\rm mono}}{\varphi (1-x)},~~
 \frac{h(m_1)}{h(1)} =
 \frac{\varphi_K^{\rm mono}}{\varphi \left[x+2(1-x)e^{-R/2}\right]}.\label{111509_1Dec15}
 \end{align}
$\varphi_K$ for large particles is obtained as
the 1RSB solution by setting $m=1$ in the first equation of
Eq.~(\ref{111509_1Dec15}) as
\begin{align}
\varphi_K^{\rm 1RSB}(x) &= \frac{\varphi_K^{\rm mono}}{1-x}.
\label{1rsb1}
\end{align}
Similarly, $\varphi_K$ for small particles is obtained as the 2RSB
solution by setting $m_1=1$ in the second equation of
Eq.~(\ref{111509_1Dec15});
\begin{align}
\varphi_K^{\rm 2RSB}(x)
& = \frac{\varphi_K^{\rm mono}}{x+2(1-x)e^{-R/2}}.\label{112115_1Dec15}
\end{align}
In the case of $m=m_1<1$, on the other hand, the glass phase is
described by the 1RSB free energy (see Sec.~\ref{121359_24Oct20}). The
1RSB and 2RSB free energies become identical when the fraction is
\begin{align}
x_c
=\frac{1-2e^{-R/2}}{2(1-e^{-R/2})}. 
\end{align}
This equation determines the phase boundary between the 1RSB and 2RSB
glass phases. When $x<x_c$, the 2RSB phase is more stable than 1RSB
phase and vice versa for $x>x_c$. For $x_c$ to be positive, $R$ must be
larger than $R_c=2\log 2$, which is the necessary condition for the 2RSB
phase or equivalently the decoupling of the two glass transitions. Note that all the
arguments above are independent of the temperature and shape of the
potential $v_{\mu\nu}(r)$, as it is obvious by rescaling the density by
\begin{align}
\hvar=\frac{\varphi}{\varphi_K^{\rm mono}}. 
\end{align}

Combining all results discussed above, we draw the thermodynamic glass
phase diagram. If $R < R_c$, the phase diagram is determined by the
1RSB-RLT and only a single glass phase exists. If $R>R_c$, four
different phases emerge as shown in FIG.~\ref{fig_4}. At very low
densities, the system is in the RS (fluid) phase where the solution with
$m=m_1=1$ is the most stable.  If $\varphi$ is large and $x$ is close to
$1$, the solution with $m=m_1<1$ is the most stable, and the system is
in the 1RSB phase where all particles are frozen. We refer to this phase
as the 1RSB \textit{double} glass phase. In this phase, the majority is
small particles, and they drive the system into the glass phase.  In
other words, large particles are embedded in vitrified small
particles. Indeed, $\varphi_K(x)$ smoothly converges to $\varphi_K^{\rm
mono}$ in the one-component limit, $x \to 1$. As $x$ decreases and
crosses $x_c$, the system undergoes the transition from the 1RSB
($m=m_1<1$) to 2RSB ($m<m_1<1$). We refer to this phase as the 2RSB
\textit{double} glass.  As $x$ decreases further, $m<m_1=1$ becomes
stable and small particles melt into a fluid phase whereas large
particles remain frozen.  We refer to this phase as the 1RSB
\textit{single} glass. The difference between the 1RSB and 2RSB double
glass phases should be emphasized.

\begin{figure}[tb]
\includegraphics[width=0.83\columnwidth]{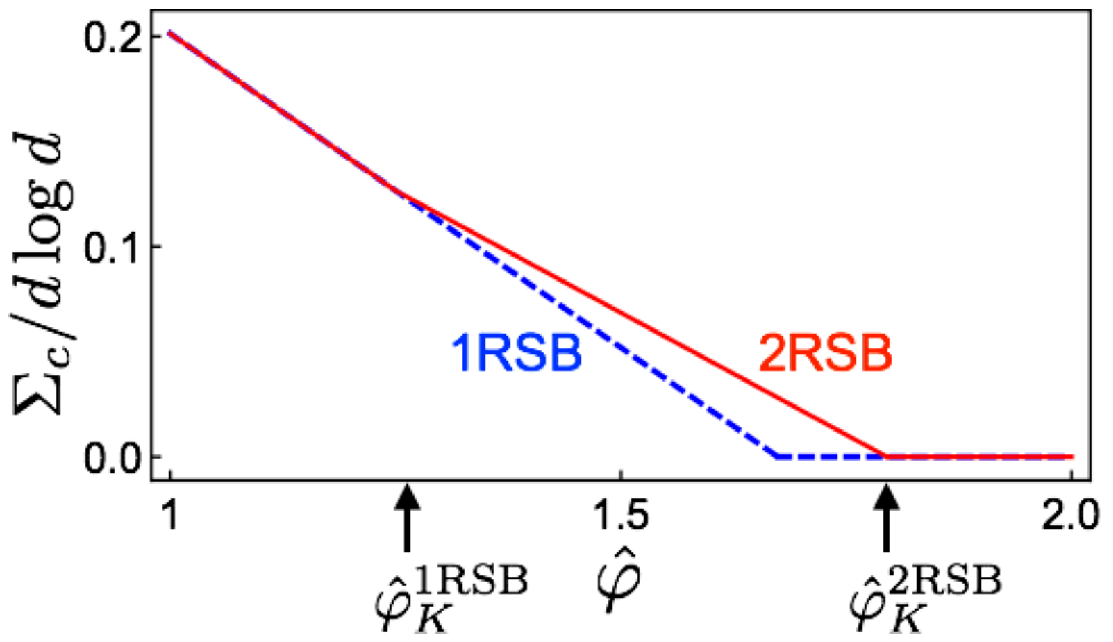} \caption{\small The
 density dependence of the configurational entropy for $R=3$ and $x=0.2$
 which is smaller than $x_c<0.36$.  The solid and dashed lines represent
 the 2RSB and 1RSB results, respectively.  The arrows indicates the
 transition point from the fluid to 1RSB(2) and the transition point
 from the 1RSB(2) to 2RSB phases.  }  \label{fig_5}
\end{figure}

As the density is increased for a fixed $x$ below $x_c$, the system
undergoes the two step glass transitions: first from the fluid to the
1RSB single glass and then to the 2RSB double glass phase. In order to
clarify the nature of this multiple transitions, we calculate the
configurational entropy, $\Sigma$, from the 2RSB free energy given by
Eq.~(\ref{160118_22Feb16}). It can be written as a sum of the two
contribution, $\Sigma= \Sigma_1 +\Sigma_2$~\cite{obuchi2010}.  Here
$\Sigma_1$ is the configurational entropy of large particles
corresponding to the large metabasins generated by large particles.
$\Sigma_2$ is the configurational entropy of small particles
corresponding to the basins inside the one of the metabasins.  We
evaluate $\Sigma$ using the asymptotic expression of the free energy,
Eq.~(\ref{153441_11Oct15}), and Eqs.~(\ref{120152_24Oct20}).
FIG.~\ref{fig_5} is the density dependence of $\Sigma$ for $R=3$ and
$x=0.2$.  The result of the (meta-stable) 1RSB solution is also shown
with the dashed line for a reference. One observes that $\Sigma$ bends
twice; first at $\hvar_K^{\rm 1RSB}$, where $\Sigma_1$ vanishes and the
second at $\hvar_K^{\rm 2RSB}$, where $\Sigma_2$ vanishes, and thus the
whole configurational entropy dies out.

\section{Summary and conclusions}

In summary, we developed a new formalism of the RLT for binary mixtures
of large and small particles based on the 2RSB ansatz. We determined the
glass phase diagram for a hardsphere-like fluid in infinite dimension.
The theory predicts that when the size ratio $R$ is larger than a
critical value, $R_c$, the hierarchical energy landscape emerges and the
decoupling of the glass transition of large and small particles takes
place. As a consequence, there arise several distinct glass phases: the
1RSB double glass, 1RSB single glass, and the 2RSB double glass phases.

It should be addressed that the 1RSB and 2RSB double glasses are
distinct phases with qualitative and topographical differences in their
free energy landscapes. The energy landscape in the 2RSB phase has the
two-step hierarchical structure where the two levels correspond to the
configurations of large and small particles, respectively. It is
desirable to design experimental setup or simulation method which allows
to delineate the difference of the two phases. We suspect that a
mechanical response or nonlinear rheology measurement would be one of
ideal
candidates~\cite{shikata2001,pham2006,koumakis2011,yoshino2014}. For
example, an anomalous two-step yielding in colloidal binary mixtures has
been reported~\cite{sentjabrskaja2013}, which may be a reflection of
complex and hierarchical energy landscape. Note that in a recent
simulation study of hard-spheres near the jamming transition, two glass
phases characterized by different status of the replica symmetry
breaking are indeed separated well by rheological
measurement~\cite{Jin2017}. The shape of the phase diagram predicted by
our theory is qualitatively consistent with experiment and numerical
results~\cite{imhof1995PRL,imhof1995,moreno2006,moreno2006jcp,hendricks2015};
the single glass phase is located at high density and small $x$ region
while the double glass phase is at high density and large $x$
region. For more quantitative comparison, it is necessary to extend our
theory to finite dimensions.


The relationship of our theory with the MCT, on the other hand, remains
somewhat elusive. First, our theory does not distinguish the self and
collective density correlations, in contrast to the MCT in finite
$d$. This simplifies the analysis but simultaneously reduces the
diversity of the phase diagram. Second, the MCT is usually identified
with the dynamical transition point of the replica liquid
theory~\footnote{ In the mean-field p-spin spherical model, the MCT is
unambiguously the dynamical counterpart of the replica
theory~\cite{castellani2005spin}. In the infinite dimensional
monodisperse particles, the dynamical transition predicted by the
replica theory share the common features with the one predicted by the
MCT~\cite{parisi2010,Maimbourg2016}.}. However our theory shows that the
description of the single and double glass phases at smaller $x$
requires the 2RSB ansatz. This means that the nature of the decoupling
predicted by the MCT is essentially different from those of
thermodynamic theory.

We believe that our higher order replica symmetric breaking picture is
not restricted to binary mixtures of disparate size ratios, but can be
adapted for other decoupling phenomena of the glass transitions, {\it
e.g.}, the decoupling of translational and rotational motions in
anisotropic particles. These are left for the future work.


\begin{acknowledgments}
 We thank G. Biroli, P. Urbani, F. Zamponi and K. Hukushima for kind
 discussions.  We acknowledge KAKENHI No.  25103005, 
 25000002, 
 and the JSPS Core-to-Core program.  H. I. was supported by Program for
 Leading Graduate Schools ``Integrative Graduate Education and Research
 in Green Natural Sciences'', MEXT, Japan.
\end{acknowledgments}

\bibliography{apssamp}

\begin{thebibliography}{61}%
\makeatletter
\providecommand \@ifxundefined [1]{%
 \@ifx{#1\undefined}
}%
\providecommand \@ifnum [1]{%
 \ifnum #1\expandafter \@firstoftwo
 \else \expandafter \@secondoftwo
 \fi
}%
\providecommand \@ifx [1]{%
 \ifx #1\expandafter \@firstoftwo
 \else \expandafter \@secondoftwo
 \fi
}%
\providecommand \natexlab [1]{#1}%
\providecommand \enquote  [1]{``#1''}%
\providecommand \bibnamefont  [1]{#1}%
\providecommand \bibfnamefont [1]{#1}%
\providecommand \citenamefont [1]{#1}%
\providecommand \href@noop [0]{\@secondoftwo}%
\providecommand \href [0]{\begingroup \@sanitize@url \@href}%
\providecommand \@href[1]{\@@startlink{#1}\@@href}%
\providecommand \@@href[1]{\endgroup#1\@@endlink}%
\providecommand \@sanitize@url [0]{\catcode `\\12\catcode `\$12\catcode
  `\&12\catcode `\#12\catcode `\^12\catcode `\_12\catcode `\%12\relax}%
\providecommand \@@startlink[1]{}%
\providecommand \@@endlink[0]{}%
\providecommand \url  [0]{\begingroup\@sanitize@url \@url }%
\providecommand \@url [1]{\endgroup\@href {#1}{\urlprefix }}%
\providecommand \urlprefix  [0]{URL }%
\providecommand \Eprint [0]{\href }%
\providecommand \doibase [0]{http://dx.doi.org/}%
\providecommand \selectlanguage [0]{\@gobble}%
\providecommand \bibinfo  [0]{\@secondoftwo}%
\providecommand \bibfield  [0]{\@secondoftwo}%
\providecommand \translation [1]{[#1]}%
\providecommand \BibitemOpen [0]{}%
\providecommand \bibitemStop [0]{}%
\providecommand \bibitemNoStop [0]{.\EOS\space}%
\providecommand \EOS [0]{\spacefactor3000\relax}%
\providecommand \BibitemShut  [1]{\csname bibitem#1\endcsname}%
\let\auto@bib@innerbib\@empty
\bibitem [{\citenamefont {Foffi}\ \emph {et~al.}(2004)\citenamefont {Foffi},
  \citenamefont {G\"otze}, \citenamefont {Sciortino}, \citenamefont
  {Tartaglia},\ and\ \citenamefont {Voigtmann}}]{Foffi2004}%
  \BibitemOpen
  \bibfield  {author} {\bibinfo {author} {\bibfnamefont {G.}~\bibnamefont
  {Foffi}}, \bibinfo {author} {\bibfnamefont {W.}~\bibnamefont {G\"otze}},
  \bibinfo {author} {\bibfnamefont {F.}~\bibnamefont {Sciortino}}, \bibinfo
  {author} {\bibfnamefont {P.}~\bibnamefont {Tartaglia}}, \ and\ \bibinfo
  {author} {\bibfnamefont {T.}~\bibnamefont {Voigtmann}},\ }\href {\doibase
  10.1103/PhysRevE.69.011505} {\bibfield  {journal} {\bibinfo  {journal} {Phys.
  Rev. E}\ }\textbf {\bibinfo {volume} {69}},\ \bibinfo {pages} {011505}
  (\bibinfo {year} {2004})}\BibitemShut {NoStop}%
\bibitem [{\citenamefont {Tanaka}\ \emph {et~al.}(2010)\citenamefont {Tanaka},
  \citenamefont {Kawasaki}, \citenamefont {Shintani},\ and\ \citenamefont
  {Watanabe}}]{Tanaka2010}%
  \BibitemOpen
  \bibfield  {author} {\bibinfo {author} {\bibfnamefont {H.}~\bibnamefont
  {Tanaka}}, \bibinfo {author} {\bibfnamefont {T.}~\bibnamefont {Kawasaki}},
  \bibinfo {author} {\bibfnamefont {H.}~\bibnamefont {Shintani}}, \ and\
  \bibinfo {author} {\bibfnamefont {K.}~\bibnamefont {Watanabe}},\ }\href
  {http://dx.doi.org/10.1038/nmat2634} {\bibfield  {journal} {\bibinfo
  {journal} {Nat Mater}\ }\textbf {\bibinfo {volume} {9}},\ \bibinfo {pages}
  {324} (\bibinfo {year} {2010})}\BibitemShut {NoStop}%
\bibitem [{\citenamefont {Angell}\ \emph {et~al.}(2000)\citenamefont {Angell},
  \citenamefont {Ngai}, \citenamefont {McKenna}, \citenamefont {McMillan},\
  and\ \citenamefont {Martin}}]{Angell2000}%
  \BibitemOpen
  \bibfield  {author} {\bibinfo {author} {\bibfnamefont {C.~A.}\ \bibnamefont
  {Angell}}, \bibinfo {author} {\bibfnamefont {K.~L.}\ \bibnamefont {Ngai}},
  \bibinfo {author} {\bibfnamefont {G.~B.}\ \bibnamefont {McKenna}}, \bibinfo
  {author} {\bibfnamefont {P.~F.}\ \bibnamefont {McMillan}}, \ and\ \bibinfo
  {author} {\bibfnamefont {S.~W.}\ \bibnamefont {Martin}},\ }\href@noop {}
  {\bibfield  {journal} {\bibinfo  {journal} {Journal of Applied Physics}\ }
  (\bibinfo {year} {2000})}\BibitemShut {NoStop}%
\bibitem [{\citenamefont {Martin}(1991)}]{martin1991}%
  \BibitemOpen
  \bibfield  {author} {\bibinfo {author} {\bibfnamefont {S.~W.}\ \bibnamefont
  {Martin}},\ }\href@noop {} {\bibfield  {journal} {\bibinfo  {journal} {J. Am.
  Soc.}\ }\textbf {\bibinfo {volume} {74}},\ \bibinfo {pages} {1767} (\bibinfo
  {year} {1991})}\BibitemShut {NoStop}%
\bibitem [{\citenamefont {Faupel}\ \emph {et~al.}(2003)\citenamefont {Faupel},
  \citenamefont {Frank}, \citenamefont {Macht}, \citenamefont {Mehrer},
  \citenamefont {Naundorf}, \citenamefont {R\"atzke}, \citenamefont {Schober},
  \citenamefont {Sharma},\ and\ \citenamefont {Teichler}}]{franz2003}%
  \BibitemOpen
  \bibfield  {author} {\bibinfo {author} {\bibfnamefont {F.}~\bibnamefont
  {Faupel}}, \bibinfo {author} {\bibfnamefont {W.}~\bibnamefont {Frank}},
  \bibinfo {author} {\bibfnamefont {M.-P.}\ \bibnamefont {Macht}}, \bibinfo
  {author} {\bibfnamefont {H.}~\bibnamefont {Mehrer}}, \bibinfo {author}
  {\bibfnamefont {V.}~\bibnamefont {Naundorf}}, \bibinfo {author}
  {\bibfnamefont {K.}~\bibnamefont {R\"atzke}}, \bibinfo {author}
  {\bibfnamefont {H.~R.}\ \bibnamefont {Schober}}, \bibinfo {author}
  {\bibfnamefont {S.~K.}\ \bibnamefont {Sharma}}, \ and\ \bibinfo {author}
  {\bibfnamefont {H.}~\bibnamefont {Teichler}},\ }\href {\doibase
  10.1103/RevModPhys.75.237} {\bibfield  {journal} {\bibinfo  {journal} {Rev.
  Mod. Phys.}\ }\textbf {\bibinfo {volume} {75}},\ \bibinfo {pages} {237}
  (\bibinfo {year} {2003})}\BibitemShut {NoStop}%
\bibitem [{\citenamefont {Mayer}\ \emph
  {et~al.}(2008{\natexlab{a}})\citenamefont {Mayer}, \citenamefont
  {Zaccarelli}, \citenamefont {Stiakakis}, \citenamefont {Likos}, \citenamefont
  {Sciortino}, \citenamefont {Munam}, \citenamefont {Gauthier}, \citenamefont
  {Hadjichristidis}, \citenamefont {Iatrou}, \citenamefont {Tartaglia} \emph
  {et~al.}}]{mayer2008}%
  \BibitemOpen
  \bibfield  {author} {\bibinfo {author} {\bibfnamefont {C.}~\bibnamefont
  {Mayer}}, \bibinfo {author} {\bibfnamefont {E.}~\bibnamefont {Zaccarelli}},
  \bibinfo {author} {\bibfnamefont {E.}~\bibnamefont {Stiakakis}}, \bibinfo
  {author} {\bibfnamefont {C.}~\bibnamefont {Likos}}, \bibinfo {author}
  {\bibfnamefont {F.}~\bibnamefont {Sciortino}}, \bibinfo {author}
  {\bibfnamefont {A.}~\bibnamefont {Munam}}, \bibinfo {author} {\bibfnamefont
  {M.}~\bibnamefont {Gauthier}}, \bibinfo {author} {\bibfnamefont
  {N.}~\bibnamefont {Hadjichristidis}}, \bibinfo {author} {\bibfnamefont
  {H.}~\bibnamefont {Iatrou}}, \bibinfo {author} {\bibfnamefont
  {P.}~\bibnamefont {Tartaglia}},  \emph {et~al.},\ }\href@noop {} {\bibfield
  {journal} {\bibinfo  {journal} {Nat. Mater.}\ }\textbf {\bibinfo {volume}
  {7}},\ \bibinfo {pages} {780} (\bibinfo {year}
  {2008}{\natexlab{a}})}\BibitemShut {NoStop}%
\bibitem [{\citenamefont {Mayer}\ \emph
  {et~al.}(2008{\natexlab{b}})\citenamefont {Mayer}, \citenamefont {Sciortino},
  \citenamefont {Likos}, \citenamefont {Tartaglia}, \citenamefont
  {L{\"{o}}wen},\ and\ \citenamefont {Zaccarelli}}]{mayer2008m}%
  \BibitemOpen
  \bibfield  {author} {\bibinfo {author} {\bibfnamefont {C.}~\bibnamefont
  {Mayer}}, \bibinfo {author} {\bibfnamefont {F.}~\bibnamefont {Sciortino}},
  \bibinfo {author} {\bibfnamefont {C.~N.}\ \bibnamefont {Likos}}, \bibinfo
  {author} {\bibfnamefont {P.}~\bibnamefont {Tartaglia}}, \bibinfo {author}
  {\bibfnamefont {H.}~\bibnamefont {L{\"{o}}wen}}, \ and\ \bibinfo {author}
  {\bibfnamefont {E.}~\bibnamefont {Zaccarelli}},\ }\href@noop {} {\bibfield
  {journal} {\bibinfo  {journal} {Macromolecules}\ }\textbf {\bibinfo {volume}
  {42}},\ \bibinfo {pages} {423} (\bibinfo {year}
  {2008}{\natexlab{b}})}\BibitemShut {NoStop}%
\bibitem [{\citenamefont {Imhof}\ and\ \citenamefont
  {Dhont}(1995{\natexlab{a}})}]{imhof1995PRL}%
  \BibitemOpen
  \bibfield  {author} {\bibinfo {author} {\bibfnamefont {A.}~\bibnamefont
  {Imhof}}\ and\ \bibinfo {author} {\bibfnamefont {J.~K.~G.}\ \bibnamefont
  {Dhont}},\ }\href {\doibase 10.1103/PhysRevLett.75.1662} {\bibfield
  {journal} {\bibinfo  {journal} {Phys. Rev. Lett.}\ }\textbf {\bibinfo
  {volume} {75}},\ \bibinfo {pages} {1662} (\bibinfo {year}
  {1995}{\natexlab{a}})}\BibitemShut {NoStop}%
\bibitem [{\citenamefont {Imhof}\ and\ \citenamefont
  {Dhont}(1995{\natexlab{b}})}]{imhof1995}%
  \BibitemOpen
  \bibfield  {author} {\bibinfo {author} {\bibfnamefont {A.}~\bibnamefont
  {Imhof}}\ and\ \bibinfo {author} {\bibfnamefont {J.~K.~G.}\ \bibnamefont
  {Dhont}},\ }\href@noop {} {\bibfield  {journal} {\bibinfo  {journal} {Phys.
  Rev. E.}\ }\textbf {\bibinfo {volume} {52}},\ \bibinfo {pages} {6345}
  (\bibinfo {year} {1995}{\natexlab{b}})}\BibitemShut {NoStop}%
\bibitem [{\citenamefont {Hendricks}\ \emph {et~al.}(2015)\citenamefont
  {Hendricks}, \citenamefont {Capellmann}, \citenamefont {Schofield},
  \citenamefont {Egelhaaf},\ and\ \citenamefont {Laurati}}]{hendricks2015}%
  \BibitemOpen
  \bibfield  {author} {\bibinfo {author} {\bibfnamefont {J.}~\bibnamefont
  {Hendricks}}, \bibinfo {author} {\bibfnamefont {R.}~\bibnamefont
  {Capellmann}}, \bibinfo {author} {\bibfnamefont {A.}~\bibnamefont
  {Schofield}}, \bibinfo {author} {\bibfnamefont {S.}~\bibnamefont {Egelhaaf}},
  \ and\ \bibinfo {author} {\bibfnamefont {M.}~\bibnamefont {Laurati}},\
  }\href@noop {} {\bibfield  {journal} {\bibinfo  {journal} {Phys. Rev. E}\
  }\textbf {\bibinfo {volume} {91}},\ \bibinfo {pages} {032308} (\bibinfo
  {year} {2015})}\BibitemShut {NoStop}%
\bibitem [{Note1()}]{Note1}%
  \BibitemOpen
  \bibinfo {note} {We use a term ``decoupling'' to mean the decoupling of the
  glass transition of small and large particles, and not necessarily mean the
  decoupling of the self and collective correlation functions. As shown later,
  the glass phase in the infinite dimensional system is characterized solely by
  the self correlation functions, not by the collective correlation
  functions.}\BibitemShut {Stop}%
\bibitem [{\citenamefont {Moreno}\ and\ \citenamefont
  {Colmenero}(2006{\natexlab{a}})}]{moreno2006}%
  \BibitemOpen
  \bibfield  {author} {\bibinfo {author} {\bibfnamefont {A.~J.}\ \bibnamefont
  {Moreno}}\ and\ \bibinfo {author} {\bibfnamefont {J.}~\bibnamefont
  {Colmenero}},\ }\href@noop {} {\bibfield  {journal} {\bibinfo  {journal}
  {Phys. Rev. E.}\ }\textbf {\bibinfo {volume} {74}},\ \bibinfo {pages}
  {021409} (\bibinfo {year} {2006}{\natexlab{a}})}\BibitemShut {NoStop}%
\bibitem [{\citenamefont {Moreno}\ and\ \citenamefont
  {Colmenero}(2006{\natexlab{b}})}]{moreno2006jcp}%
  \BibitemOpen
  \bibfield  {author} {\bibinfo {author} {\bibfnamefont {A.~J.}\ \bibnamefont
  {Moreno}}\ and\ \bibinfo {author} {\bibfnamefont {J.}~\bibnamefont
  {Colmenero}},\ }\href@noop {} {\bibfield  {journal} {\bibinfo  {journal} {The
  Journal of chemical physics}\ }\textbf {\bibinfo {volume} {125}},\ \bibinfo
  {pages} {164507} (\bibinfo {year} {2006}{\natexlab{b}})}\BibitemShut
  {NoStop}%
\bibitem [{\citenamefont {Voigtmann}\ and\ \citenamefont
  {Horbach}(2009)}]{voigtmann2009}%
  \BibitemOpen
  \bibfield  {author} {\bibinfo {author} {\bibfnamefont {T.}~\bibnamefont
  {Voigtmann}}\ and\ \bibinfo {author} {\bibfnamefont {J.}~\bibnamefont
  {Horbach}},\ }\href {\doibase 10.1103/PhysRevLett.103.205901} {\bibfield
  {journal} {\bibinfo  {journal} {Phys. Rev. Lett.}\ }\textbf {\bibinfo
  {volume} {103}},\ \bibinfo {pages} {205901} (\bibinfo {year}
  {2009})}\BibitemShut {NoStop}%
\bibitem [{\citenamefont {Bengtzelius}\ \emph {et~al.}(1984)\citenamefont
  {Bengtzelius}, \citenamefont {Gotze},\ and\ \citenamefont
  {Sjolander}}]{bengtzelius1984}%
  \BibitemOpen
  \bibfield  {author} {\bibinfo {author} {\bibfnamefont {U.}~\bibnamefont
  {Bengtzelius}}, \bibinfo {author} {\bibfnamefont {W.}~\bibnamefont {Gotze}},
  \ and\ \bibinfo {author} {\bibfnamefont {A.}~\bibnamefont {Sjolander}},\
  }\href@noop {} {\bibfield  {journal} {\bibinfo  {journal} {J. Phys. C}\
  }\textbf {\bibinfo {volume} {17}},\ \bibinfo {pages} {5915} (\bibinfo {year}
  {1984})}\BibitemShut {NoStop}%
\bibitem [{\citenamefont {Gotze}(2009)}]{gotze2009}%
  \BibitemOpen
  \bibfield  {author} {\bibinfo {author} {\bibfnamefont {W.}~\bibnamefont
  {Gotze}},\ }\href@noop {} {\emph {\bibinfo {title} {Complex dynamics of
  glass-forming liquids}}}\ (\bibinfo  {publisher} {Oxford University Press},\
  \bibinfo {year} {2009})\BibitemShut {NoStop}%
\bibitem [{\citenamefont {Bosse}\ and\ \citenamefont
  {Thakur}(1987)}]{bosse1987}%
  \BibitemOpen
  \bibfield  {author} {\bibinfo {author} {\bibfnamefont {J.}~\bibnamefont
  {Bosse}}\ and\ \bibinfo {author} {\bibfnamefont {J.~S.}\ \bibnamefont
  {Thakur}},\ }\href@noop {} {\bibfield  {journal} {\bibinfo  {journal} {Phys.
  Rev. Lett.}\ }\textbf {\bibinfo {volume} {59}},\ \bibinfo {pages} {998}
  (\bibinfo {year} {1987})}\BibitemShut {NoStop}%
\bibitem [{\citenamefont {Bosse}\ and\ \citenamefont
  {Kaneko}(1995)}]{bosse1995}%
  \BibitemOpen
  \bibfield  {author} {\bibinfo {author} {\bibfnamefont {J.}~\bibnamefont
  {Bosse}}\ and\ \bibinfo {author} {\bibfnamefont {Y.}~\bibnamefont {Kaneko}},\
  }\href@noop {} {\bibfield  {journal} {\bibinfo  {journal} {Phys. Rev. Lett.}\
  }\textbf {\bibinfo {volume} {74}},\ \bibinfo {pages} {4023} (\bibinfo {year}
  {1995})}\BibitemShut {NoStop}%
\bibitem [{\citenamefont {Voigtmann}(2011)}]{voigtmann2011}%
  \BibitemOpen
  \bibfield  {author} {\bibinfo {author} {\bibfnamefont {T.}~\bibnamefont
  {Voigtmann}},\ }\href@noop {} {\bibfield  {journal} {\bibinfo  {journal} {EPL
  (Europhysics Letters)}\ }\textbf {\bibinfo {volume} {96}},\ \bibinfo {pages}
  {36006} (\bibinfo {year} {2011})}\BibitemShut {NoStop}%
\bibitem [{\citenamefont {L\'azaro-L\'azaro}\ \emph {et~al.}(2019)\citenamefont
  {L\'azaro-L\'azaro}, \citenamefont {Perera-Burgos}, \citenamefont {Laermann},
  \citenamefont {Sentjabrskaja}, \citenamefont {P\'erez-\'Angel}, \citenamefont
  {Laurati}, \citenamefont {Egelhaaf}, \citenamefont {Medina-Noyola},
  \citenamefont {Voigtmann}, \citenamefont {Casta\~neda Priego},\ and\
  \citenamefont {Elizondo-Aguilera}}]{PhysRevE.99.042603}%
  \BibitemOpen
  \bibfield  {author} {\bibinfo {author} {\bibfnamefont {E.}~\bibnamefont
  {L\'azaro-L\'azaro}}, \bibinfo {author} {\bibfnamefont {J.~A.}\ \bibnamefont
  {Perera-Burgos}}, \bibinfo {author} {\bibfnamefont {P.}~\bibnamefont
  {Laermann}}, \bibinfo {author} {\bibfnamefont {T.}~\bibnamefont
  {Sentjabrskaja}}, \bibinfo {author} {\bibfnamefont {G.}~\bibnamefont
  {P\'erez-\'Angel}}, \bibinfo {author} {\bibfnamefont {M.}~\bibnamefont
  {Laurati}}, \bibinfo {author} {\bibfnamefont {S.~U.}\ \bibnamefont
  {Egelhaaf}}, \bibinfo {author} {\bibfnamefont {M.}~\bibnamefont
  {Medina-Noyola}}, \bibinfo {author} {\bibfnamefont {T.}~\bibnamefont
  {Voigtmann}}, \bibinfo {author} {\bibfnamefont {R.}~\bibnamefont {Casta\~neda
  Priego}}, \ and\ \bibinfo {author} {\bibfnamefont {L.~F.}\ \bibnamefont
  {Elizondo-Aguilera}},\ }\href {\doibase 10.1103/PhysRevE.99.042603}
  {\bibfield  {journal} {\bibinfo  {journal} {Phys. Rev. E}\ }\textbf {\bibinfo
  {volume} {99}},\ \bibinfo {pages} {042603} (\bibinfo {year}
  {2019})}\BibitemShut {NoStop}%
\bibitem [{\citenamefont {Elizondo-Aguilera}\ and\ \citenamefont
  {Voigtmann}(2019)}]{PhysRevE.100.042601}%
  \BibitemOpen
  \bibfield  {author} {\bibinfo {author} {\bibfnamefont {L.~F.}\ \bibnamefont
  {Elizondo-Aguilera}}\ and\ \bibinfo {author} {\bibfnamefont {T.}~\bibnamefont
  {Voigtmann}},\ }\href {\doibase 10.1103/PhysRevE.100.042601} {\bibfield
  {journal} {\bibinfo  {journal} {Phys. Rev. E}\ }\textbf {\bibinfo {volume}
  {100}},\ \bibinfo {pages} {042601} (\bibinfo {year} {2019})}\BibitemShut
  {NoStop}%
\bibitem [{\citenamefont {Kirkpatrick}\ and\ \citenamefont
  {Wolynes}(1987{\natexlab{a}})}]{Kirkpatrick1987b}%
  \BibitemOpen
  \bibfield  {author} {\bibinfo {author} {\bibfnamefont {T.~R.}\ \bibnamefont
  {Kirkpatrick}}\ and\ \bibinfo {author} {\bibfnamefont {P.~G.}\ \bibnamefont
  {Wolynes}},\ }\href {\doibase 10.1103/PhysRevA.35.3072} {\bibfield  {journal}
  {\bibinfo  {journal} {Phys. Rev. A}\ }\textbf {\bibinfo {volume} {35}},\
  \bibinfo {pages} {3072} (\bibinfo {year} {1987}{\natexlab{a}})}\BibitemShut
  {NoStop}%
\bibitem [{\citenamefont {Schmid}\ and\ \citenamefont
  {Schilling}(2010)}]{Schmid2010}%
  \BibitemOpen
  \bibfield  {author} {\bibinfo {author} {\bibfnamefont {B.}~\bibnamefont
  {Schmid}}\ and\ \bibinfo {author} {\bibfnamefont {R.}~\bibnamefont
  {Schilling}},\ }\href {\doibase 10.1103/PhysRevE.81.041502} {\bibfield
  {journal} {\bibinfo  {journal} {Phys. Rev. E}\ }\textbf {\bibinfo {volume}
  {81}},\ \bibinfo {pages} {041502} (\bibinfo {year} {2010})}\BibitemShut
  {NoStop}%
\bibitem [{\citenamefont {Ikeda}\ and\ \citenamefont
  {Miyazaki}(2010)}]{Ikeda2010}%
  \BibitemOpen
  \bibfield  {author} {\bibinfo {author} {\bibfnamefont {A.}~\bibnamefont
  {Ikeda}}\ and\ \bibinfo {author} {\bibfnamefont {K.}~\bibnamefont
  {Miyazaki}},\ }\href {\doibase 10.1103/PhysRevLett.104.255704} {\bibfield
  {journal} {\bibinfo  {journal} {Phys. Rev. Lett.}\ }\textbf {\bibinfo
  {volume} {104}},\ \bibinfo {pages} {255704} (\bibinfo {year}
  {2010})}\BibitemShut {NoStop}%
\bibitem [{\citenamefont {Jin}\ and\ \citenamefont
  {Charbonneau}(2015)}]{Jin2015}%
  \BibitemOpen
  \bibfield  {author} {\bibinfo {author} {\bibfnamefont {Y.}~\bibnamefont
  {Jin}}\ and\ \bibinfo {author} {\bibfnamefont {P.}~\bibnamefont
  {Charbonneau}},\ }\href {\doibase 10.1103/PhysRevE.91.042313} {\bibfield
  {journal} {\bibinfo  {journal} {Phys. Rev. E}\ }\textbf {\bibinfo {volume}
  {91}},\ \bibinfo {pages} {042313} (\bibinfo {year} {2015})}\BibitemShut
  {NoStop}%
\bibitem [{\citenamefont {Maimbourg}\ \emph {et~al.}(2016)\citenamefont
  {Maimbourg}, \citenamefont {Kurchan},\ and\ \citenamefont
  {Zamponi}}]{Maimbourg2016}%
  \BibitemOpen
  \bibfield  {author} {\bibinfo {author} {\bibfnamefont {T.}~\bibnamefont
  {Maimbourg}}, \bibinfo {author} {\bibfnamefont {J.}~\bibnamefont {Kurchan}},
  \ and\ \bibinfo {author} {\bibfnamefont {F.}~\bibnamefont {Zamponi}},\ }\href
  {\doibase 10.1103/PhysRevLett.116.015902} {\bibfield  {journal} {\bibinfo
  {journal} {Phys. Rev. Lett.}\ }\textbf {\bibinfo {volume} {116}},\ \bibinfo
  {pages} {015902} (\bibinfo {year} {2016})}\BibitemShut {NoStop}%
\bibitem [{\citenamefont {Monasson}(1995)}]{monasson1995}%
  \BibitemOpen
  \bibfield  {author} {\bibinfo {author} {\bibfnamefont {R.}~\bibnamefont
  {Monasson}},\ }\href@noop {} {\bibfield  {journal} {\bibinfo  {journal}
  {Phys. Rev. Lett.}\ }\textbf {\bibinfo {volume} {75}},\ \bibinfo {pages}
  {2847} (\bibinfo {year} {1995})}\BibitemShut {NoStop}%
\bibitem [{\citenamefont {M\'ezard}\ and\ \citenamefont
  {Parisi}(1999)}]{mezard1999}%
  \BibitemOpen
  \bibfield  {author} {\bibinfo {author} {\bibfnamefont {M.}~\bibnamefont
  {M\'ezard}}\ and\ \bibinfo {author} {\bibfnamefont {G.}~\bibnamefont
  {Parisi}},\ }\href {\doibase 10.1103/PhysRevLett.82.747} {\bibfield
  {journal} {\bibinfo  {journal} {Phys. Rev. Lett.}\ }\textbf {\bibinfo
  {volume} {82}},\ \bibinfo {pages} {747} (\bibinfo {year} {1999})}\BibitemShut
  {NoStop}%
\bibitem [{\citenamefont {Parisi}\ and\ \citenamefont
  {Zamponi}(2010)}]{parisi2010}%
  \BibitemOpen
  \bibfield  {author} {\bibinfo {author} {\bibfnamefont {G.}~\bibnamefont
  {Parisi}}\ and\ \bibinfo {author} {\bibfnamefont {F.}~\bibnamefont
  {Zamponi}},\ }\href@noop {} {\bibfield  {journal} {\bibinfo  {journal} {Rev.
  Mod. Phys.}\ }\textbf {\bibinfo {volume} {82}},\ \bibinfo {pages} {789}
  (\bibinfo {year} {2010})}\BibitemShut {NoStop}%
\bibitem [{\citenamefont {Kirkpatrick}\ and\ \citenamefont
  {Wolynes}(1987{\natexlab{b}})}]{PhysRevB.36.8552}%
  \BibitemOpen
  \bibfield  {author} {\bibinfo {author} {\bibfnamefont {T.~R.}\ \bibnamefont
  {Kirkpatrick}}\ and\ \bibinfo {author} {\bibfnamefont {P.~G.}\ \bibnamefont
  {Wolynes}},\ }\href {\doibase 10.1103/PhysRevB.36.8552} {\bibfield  {journal}
  {\bibinfo  {journal} {Phys. Rev. B}\ }\textbf {\bibinfo {volume} {36}},\
  \bibinfo {pages} {8552} (\bibinfo {year} {1987}{\natexlab{b}})}\BibitemShut
  {NoStop}%
\bibitem [{\citenamefont {Kirkpatrick}\ \emph {et~al.}(1989)\citenamefont
  {Kirkpatrick}, \citenamefont {Thirumalai},\ and\ \citenamefont
  {Wolynes}}]{kirkpatrick1989}%
  \BibitemOpen
  \bibfield  {author} {\bibinfo {author} {\bibfnamefont {T.}~\bibnamefont
  {Kirkpatrick}}, \bibinfo {author} {\bibfnamefont {D.}~\bibnamefont
  {Thirumalai}}, \ and\ \bibinfo {author} {\bibfnamefont {P.~G.}\ \bibnamefont
  {Wolynes}},\ }\href@noop {} {\bibfield  {journal} {\bibinfo  {journal}
  {Physical Review A}\ }\textbf {\bibinfo {volume} {40}},\ \bibinfo {pages}
  {1045} (\bibinfo {year} {1989})}\BibitemShut {NoStop}%
\bibitem [{\citenamefont {Bouchaud}\ and\ \citenamefont
  {Biroli}(2004)}]{bouchaud2004}%
  \BibitemOpen
  \bibfield  {author} {\bibinfo {author} {\bibfnamefont {J.-P.}\ \bibnamefont
  {Bouchaud}}\ and\ \bibinfo {author} {\bibfnamefont {G.}~\bibnamefont
  {Biroli}},\ }\href@noop {} {\bibfield  {journal} {\bibinfo  {journal} {The
  Journal of chemical physics}\ }\textbf {\bibinfo {volume} {121}},\ \bibinfo
  {pages} {7347} (\bibinfo {year} {2004})}\BibitemShut {NoStop}%
\bibitem [{\citenamefont {Kurchan}\ \emph {et~al.}(2012)\citenamefont
  {Kurchan}, \citenamefont {Parisi},\ and\ \citenamefont
  {Zamponi}}]{Kurchan2012}%
  \BibitemOpen
  \bibfield  {author} {\bibinfo {author} {\bibfnamefont {J.}~\bibnamefont
  {Kurchan}}, \bibinfo {author} {\bibfnamefont {G.}~\bibnamefont {Parisi}}, \
  and\ \bibinfo {author} {\bibfnamefont {F.}~\bibnamefont {Zamponi}},\ }\href
  {http://stacks.iop.org/1742-5468/2012/i=10/a=P10012} {\bibfield  {journal}
  {\bibinfo  {journal} {Journal of Statistical Mechanics: Theory and
  Experiment}\ }\textbf {\bibinfo {volume} {2012}},\ \bibinfo {pages} {P10012}
  (\bibinfo {year} {2012})}\BibitemShut {NoStop}%
\bibitem [{\citenamefont {Coluzzi}\ \emph {et~al.}(1999)\citenamefont
  {Coluzzi}, \citenamefont {M{\'e}zard}, \citenamefont {Parisi},\ and\
  \citenamefont {Verrocchio}}]{coluzzi1999}%
  \BibitemOpen
  \bibfield  {author} {\bibinfo {author} {\bibfnamefont {B.}~\bibnamefont
  {Coluzzi}}, \bibinfo {author} {\bibfnamefont {M.}~\bibnamefont {M{\'e}zard}},
  \bibinfo {author} {\bibfnamefont {G.}~\bibnamefont {Parisi}}, \ and\ \bibinfo
  {author} {\bibfnamefont {P.}~\bibnamefont {Verrocchio}},\ }\href@noop {}
  {\bibfield  {journal} {\bibinfo  {journal} {J. Chem. Phys.}\ }\textbf
  {\bibinfo {volume} {111}},\ \bibinfo {pages} {9039} (\bibinfo {year}
  {1999})}\BibitemShut {NoStop}%
\bibitem [{\citenamefont {Coluzzi}\ \emph {et~al.}(2000)\citenamefont
  {Coluzzi}, \citenamefont {Parisi},\ and\ \citenamefont
  {Verrocchio}}]{coluzzi2000}%
  \BibitemOpen
  \bibfield  {author} {\bibinfo {author} {\bibfnamefont {B.}~\bibnamefont
  {Coluzzi}}, \bibinfo {author} {\bibfnamefont {G.}~\bibnamefont {Parisi}}, \
  and\ \bibinfo {author} {\bibfnamefont {P.}~\bibnamefont {Verrocchio}},\
  }\href@noop {} {\bibfield  {journal} {\bibinfo  {journal} {J. Chem. Phys.}\
  }\textbf {\bibinfo {volume} {112}},\ \bibinfo {pages} {2933} (\bibinfo {year}
  {2000})}\BibitemShut {NoStop}%
\bibitem [{\citenamefont {Biazzo}\ \emph {et~al.}(2009)\citenamefont {Biazzo},
  \citenamefont {Caltagirone}, \citenamefont {Parisi},\ and\ \citenamefont
  {Zamponi}}]{biazzo2009}%
  \BibitemOpen
  \bibfield  {author} {\bibinfo {author} {\bibfnamefont {I.}~\bibnamefont
  {Biazzo}}, \bibinfo {author} {\bibfnamefont {F.}~\bibnamefont {Caltagirone}},
  \bibinfo {author} {\bibfnamefont {G.}~\bibnamefont {Parisi}}, \ and\ \bibinfo
  {author} {\bibfnamefont {F.}~\bibnamefont {Zamponi}},\ }\href@noop {}
  {\bibfield  {journal} {\bibinfo  {journal} {Phys. Rev. Lett.}\ }\textbf
  {\bibinfo {volume} {102}},\ \bibinfo {pages} {195701} (\bibinfo {year}
  {2009})}\BibitemShut {NoStop}%
\bibitem [{\citenamefont {Biazzo}\ \emph {et~al.}(2010)\citenamefont {Biazzo},
  \citenamefont {Caltagirone}, \citenamefont {Parisi},\ and\ \citenamefont
  {Zamponi}}]{biazzo2010}%
  \BibitemOpen
  \bibfield  {author} {\bibinfo {author} {\bibfnamefont {I.}~\bibnamefont
  {Biazzo}}, \bibinfo {author} {\bibfnamefont {F.}~\bibnamefont {Caltagirone}},
  \bibinfo {author} {\bibfnamefont {G.}~\bibnamefont {Parisi}}, \ and\ \bibinfo
  {author} {\bibfnamefont {F.}~\bibnamefont {Zamponi}},\ }\href@noop {}
  {\bibfield  {journal} {\bibinfo  {journal} {J. Chem. Phys.}\ }\textbf
  {\bibinfo {volume} {132}},\ \bibinfo {pages} {176101} (\bibinfo {year}
  {2010})}\BibitemShut {NoStop}%
\bibitem [{\citenamefont {Ikeda}\ and\ \citenamefont
  {Ikeda}(2016)}]{ikeda2016}%
  \BibitemOpen
  \bibfield  {author} {\bibinfo {author} {\bibfnamefont {H.}~\bibnamefont
  {Ikeda}}\ and\ \bibinfo {author} {\bibfnamefont {A.}~\bibnamefont {Ikeda}},\
  }\href@noop {} {\bibfield  {journal} {\bibinfo  {journal} {J. Stat. Mech.
  Theor. Exp.}\ }\textbf {\bibinfo {volume} {2016}},\ \bibinfo {pages} {074006}
  (\bibinfo {year} {2016})}\BibitemShut {NoStop}%
\bibitem [{\citenamefont {Crisanti}\ and\ \citenamefont
  {Leuzzi}(2007)}]{crisanti2007}%
  \BibitemOpen
  \bibfield  {author} {\bibinfo {author} {\bibfnamefont {A.}~\bibnamefont
  {Crisanti}}\ and\ \bibinfo {author} {\bibfnamefont {L.}~\bibnamefont
  {Leuzzi}},\ }\href@noop {} {\bibfield  {journal} {\bibinfo  {journal} {Phys.
  Rev. B}\ }\textbf {\bibinfo {volume} {76}},\ \bibinfo {pages} {184417}
  (\bibinfo {year} {2007})}\BibitemShut {NoStop}%
\bibitem [{\citenamefont {Crisanti}\ \emph {et~al.}(2011)\citenamefont
  {Crisanti}, \citenamefont {Leuzzi},\ and\ \citenamefont
  {Paoluzzi}}]{crisanti2011}%
  \BibitemOpen
  \bibfield  {author} {\bibinfo {author} {\bibfnamefont {A.}~\bibnamefont
  {Crisanti}}, \bibinfo {author} {\bibfnamefont {L.}~\bibnamefont {Leuzzi}}, \
  and\ \bibinfo {author} {\bibfnamefont {M.}~\bibnamefont {Paoluzzi}},\
  }\href@noop {} {\bibfield  {journal} {\bibinfo  {journal} {Eur. Phys. J. E}\
  }\textbf {\bibinfo {volume} {34}},\ \bibinfo {pages} {1} (\bibinfo {year}
  {2011})}\BibitemShut {NoStop}%
\bibitem [{\citenamefont {Crisanti}\ and\ \citenamefont
  {Leuzzi}(2015)}]{crisanti2015}%
  \BibitemOpen
  \bibfield  {author} {\bibinfo {author} {\bibfnamefont {A.}~\bibnamefont
  {Crisanti}}\ and\ \bibinfo {author} {\bibfnamefont {L.}~\bibnamefont
  {Leuzzi}},\ }\href@noop {} {\bibfield  {journal} {\bibinfo  {journal} {J.
  Non-Cryst. Solids}\ }\textbf {\bibinfo {volume} {407}},\ \bibinfo {pages}
  {110} (\bibinfo {year} {2015})}\BibitemShut {NoStop}%
\bibitem [{\citenamefont {Goldstein}(1969)}]{Goldstein1969}%
  \BibitemOpen
  \bibfield  {author} {\bibinfo {author} {\bibfnamefont {M.}~\bibnamefont
  {Goldstein}},\ }\href@noop {} {\bibfield  {journal} {\bibinfo  {journal} {J.
  Chem. Phys.}\ }\textbf {\bibinfo {volume} {51}},\ \bibinfo {pages} {3728}
  (\bibinfo {year} {1969})}\BibitemShut {NoStop}%
\bibitem [{\citenamefont {Sastry}\ \emph {et~al.}(1998)\citenamefont {Sastry},
  \citenamefont {Debenedetti},\ and\ \citenamefont {Stillinger}}]{Sastry1998}%
  \BibitemOpen
  \bibfield  {author} {\bibinfo {author} {\bibfnamefont {S.}~\bibnamefont
  {Sastry}}, \bibinfo {author} {\bibfnamefont {P.~G.}\ \bibnamefont
  {Debenedetti}}, \ and\ \bibinfo {author} {\bibfnamefont {F.~H.}\ \bibnamefont
  {Stillinger}},\ }\href {http://dx.doi.org/10.1038/31189} {\bibfield
  {journal} {\bibinfo  {journal} {Nature}\ }\textbf {\bibinfo {volume} {393}},\
  \bibinfo {pages} {554} (\bibinfo {year} {1998})}\BibitemShut {NoStop}%
\bibitem [{\citenamefont {Obuchi}\ \emph {et~al.}(2010)\citenamefont {Obuchi},
  \citenamefont {Takahashi},\ and\ \citenamefont {Takeda}}]{obuchi2010}%
  \BibitemOpen
  \bibfield  {author} {\bibinfo {author} {\bibfnamefont {T.}~\bibnamefont
  {Obuchi}}, \bibinfo {author} {\bibfnamefont {K.}~\bibnamefont {Takahashi}}, \
  and\ \bibinfo {author} {\bibfnamefont {K.}~\bibnamefont {Takeda}},\
  }\href@noop {} {\bibfield  {journal} {\bibinfo  {journal} {J. Phys. A}\
  }\textbf {\bibinfo {volume} {43}},\ \bibinfo {pages} {485004} (\bibinfo
  {year} {2010})}\BibitemShut {NoStop}%
\bibitem [{\citenamefont {Ikeda}\ \emph {et~al.}(2016)\citenamefont {Ikeda},
  \citenamefont {Miyazaki},\ and\ \citenamefont {Ikeda}}]{hikedanote}%
  \BibitemOpen
  \bibfield  {author} {\bibinfo {author} {\bibfnamefont {H.}~\bibnamefont
  {Ikeda}}, \bibinfo {author} {\bibfnamefont {K.}~\bibnamefont {Miyazaki}}, \
  and\ \bibinfo {author} {\bibfnamefont {A.}~\bibnamefont {Ikeda}},\ }\href
  {\doibase 10.1063/1.4969072} {\bibfield  {journal} {\bibinfo  {journal} {J.
  Chem. Phys.}\ }\textbf {\bibinfo {volume} {145}},\ \bibinfo {pages} {216101}
  (\bibinfo {year} {2016})}\BibitemShut {NoStop}%
\bibitem [{\citenamefont {Ozawa}\ and\ \citenamefont
  {Berthier}(2017)}]{ozawa2016}%
  \BibitemOpen
  \bibfield  {author} {\bibinfo {author} {\bibfnamefont {M.}~\bibnamefont
  {Ozawa}}\ and\ \bibinfo {author} {\bibfnamefont {L.}~\bibnamefont
  {Berthier}},\ }\href@noop {} {\bibfield  {journal} {\bibinfo  {journal} {J.
  Chem. Phys.}\ }\textbf {\bibinfo {volume} {146}},\ \bibinfo {pages} {014502}
  (\bibinfo {year} {2017})}\BibitemShut {NoStop}%
\bibitem [{\citenamefont {Ikeda}\ \emph {et~al.}(2017)\citenamefont {Ikeda},
  \citenamefont {Zamponi},\ and\ \citenamefont {Ikeda}}]{ikeda2017}%
  \BibitemOpen
  \bibfield  {author} {\bibinfo {author} {\bibfnamefont {H.}~\bibnamefont
  {Ikeda}}, \bibinfo {author} {\bibfnamefont {F.}~\bibnamefont {Zamponi}}, \
  and\ \bibinfo {author} {\bibfnamefont {A.}~\bibnamefont {Ikeda}},\
  }\href@noop {} {\bibfield  {journal} {\bibinfo  {journal} {The Journal of
  Chemical Physics}\ }\textbf {\bibinfo {volume} {147}},\ \bibinfo {pages}
  {234506} (\bibinfo {year} {2017})}\BibitemShut {NoStop}%
\bibitem [{\citenamefont {Ikeda}\ and\ \citenamefont
  {Zamponi}(2019)}]{ikeda2019effect}%
  \BibitemOpen
  \bibfield  {author} {\bibinfo {author} {\bibfnamefont {H.}~\bibnamefont
  {Ikeda}}\ and\ \bibinfo {author} {\bibfnamefont {F.}~\bibnamefont
  {Zamponi}},\ }\href@noop {} {\bibfield  {journal} {\bibinfo  {journal}
  {Journal of Statistical Mechanics: Theory and Experiment}\ }\textbf {\bibinfo
  {volume} {2019}},\ \bibinfo {pages} {054001} (\bibinfo {year}
  {2019})}\BibitemShut {NoStop}%
\bibitem [{\citenamefont {Parisi}\ \emph {et~al.}(2020)\citenamefont {Parisi},
  \citenamefont {Urbani},\ and\ \citenamefont {Zamponi}}]{parisi2020theory}%
  \BibitemOpen
  \bibfield  {author} {\bibinfo {author} {\bibfnamefont {G.}~\bibnamefont
  {Parisi}}, \bibinfo {author} {\bibfnamefont {P.}~\bibnamefont {Urbani}}, \
  and\ \bibinfo {author} {\bibfnamefont {F.}~\bibnamefont {Zamponi}},\
  }\href@noop {} {\emph {\bibinfo {title} {Theory of simple glasses: exact
  solutions in infinite dimensions}}}\ (\bibinfo  {publisher} {Cambridge
  University Press},\ \bibinfo {year} {2020})\BibitemShut {NoStop}%
\bibitem [{\citenamefont {Charbonneau}\ \emph {et~al.}(2014)\citenamefont
  {Charbonneau}, \citenamefont {Kurchan}, \citenamefont {Parisi}, \citenamefont
  {Urbani},\ and\ \citenamefont {Zamponi}}]{charbonneau2014}%
  \BibitemOpen
  \bibfield  {author} {\bibinfo {author} {\bibfnamefont {P.}~\bibnamefont
  {Charbonneau}}, \bibinfo {author} {\bibfnamefont {J.}~\bibnamefont
  {Kurchan}}, \bibinfo {author} {\bibfnamefont {G.}~\bibnamefont {Parisi}},
  \bibinfo {author} {\bibfnamefont {P.}~\bibnamefont {Urbani}}, \ and\ \bibinfo
  {author} {\bibfnamefont {F.}~\bibnamefont {Zamponi}},\ }\href@noop {}
  {\bibfield  {journal} {\bibinfo  {journal} {J. Stat. Mech: Theory Exp.}\
  }\textbf {\bibinfo {volume} {2014}},\ \bibinfo {pages} {P10009} (\bibinfo
  {year} {2014})}\BibitemShut {NoStop}%
\bibitem [{Note2()}]{Note2}%
  \BibitemOpen
  \bibinfo {note} {Note that $\rho _\alpha $ have only the information of the
  \protect \textit {tagged} variables and not of the \protect \textit
  {collective} variables.}\BibitemShut {Stop}%
\bibitem [{\citenamefont {Asakura}\ and\ \citenamefont
  {Oosawa}(1954)}]{asakura1954}%
  \BibitemOpen
  \bibfield  {author} {\bibinfo {author} {\bibfnamefont {S.}~\bibnamefont
  {Asakura}}\ and\ \bibinfo {author} {\bibfnamefont {F.}~\bibnamefont
  {Oosawa}},\ }\href@noop {} {\bibfield  {journal} {\bibinfo  {journal} {Chem.
  Phys.}\ ,\ \bibinfo {pages} {1255}} (\bibinfo {year} {1954})}\BibitemShut
  {NoStop}%
\bibitem [{\citenamefont {Dijkstra}\ \emph {et~al.}(1999)\citenamefont
  {Dijkstra}, \citenamefont {van Roij},\ and\ \citenamefont
  {Evans}}]{dijkstra1999}%
  \BibitemOpen
  \bibfield  {author} {\bibinfo {author} {\bibfnamefont {M.}~\bibnamefont
  {Dijkstra}}, \bibinfo {author} {\bibfnamefont {R.}~\bibnamefont {van Roij}},
  \ and\ \bibinfo {author} {\bibfnamefont {R.}~\bibnamefont {Evans}},\ }\href
  {\doibase 10.1103/PhysRevE.59.5744} {\bibfield  {journal} {\bibinfo
  {journal} {Phys. Rev. E}\ }\textbf {\bibinfo {volume} {59}},\ \bibinfo
  {pages} {5744} (\bibinfo {year} {1999})}\BibitemShut {NoStop}%
\bibitem [{\citenamefont {Shikata}(2001)}]{shikata2001}%
  \BibitemOpen
  \bibfield  {author} {\bibinfo {author} {\bibfnamefont {T.}~\bibnamefont
  {Shikata}},\ }\href@noop {} {\bibfield  {journal} {\bibinfo  {journal}
  {Chemical engineering science}\ }\textbf {\bibinfo {volume} {56}},\ \bibinfo
  {pages} {2957} (\bibinfo {year} {2001})}\BibitemShut {NoStop}%
\bibitem [{\citenamefont {Pham}\ \emph {et~al.}(2006)\citenamefont {Pham},
  \citenamefont {Petekidis}, \citenamefont {Vlassopoulos}, \citenamefont
  {Egelhaaf}, \citenamefont {Pusey},\ and\ \citenamefont {Poon}}]{pham2006}%
  \BibitemOpen
  \bibfield  {author} {\bibinfo {author} {\bibfnamefont {K.}~\bibnamefont
  {Pham}}, \bibinfo {author} {\bibfnamefont {G.}~\bibnamefont {Petekidis}},
  \bibinfo {author} {\bibfnamefont {D.}~\bibnamefont {Vlassopoulos}}, \bibinfo
  {author} {\bibfnamefont {S.}~\bibnamefont {Egelhaaf}}, \bibinfo {author}
  {\bibfnamefont {P.}~\bibnamefont {Pusey}}, \ and\ \bibinfo {author}
  {\bibfnamefont {W.}~\bibnamefont {Poon}},\ }\href@noop {} {\bibfield
  {journal} {\bibinfo  {journal} {EPL (Europhysics Letters)}\ }\textbf
  {\bibinfo {volume} {75}},\ \bibinfo {pages} {624} (\bibinfo {year}
  {2006})}\BibitemShut {NoStop}%
\bibitem [{\citenamefont {Koumakis}\ and\ \citenamefont
  {Petekidis}(2011)}]{koumakis2011}%
  \BibitemOpen
  \bibfield  {author} {\bibinfo {author} {\bibfnamefont {N.}~\bibnamefont
  {Koumakis}}\ and\ \bibinfo {author} {\bibfnamefont {G.}~\bibnamefont
  {Petekidis}},\ }\href@noop {} {\bibfield  {journal} {\bibinfo  {journal}
  {Soft Matter}\ }\textbf {\bibinfo {volume} {7}},\ \bibinfo {pages} {2456}
  (\bibinfo {year} {2011})}\BibitemShut {NoStop}%
\bibitem [{\citenamefont {Yoshino}\ and\ \citenamefont
  {Zamponi}(2014)}]{yoshino2014}%
  \BibitemOpen
  \bibfield  {author} {\bibinfo {author} {\bibfnamefont {H.}~\bibnamefont
  {Yoshino}}\ and\ \bibinfo {author} {\bibfnamefont {F.}~\bibnamefont
  {Zamponi}},\ }\href {\doibase 10.1103/PhysRevE.90.022302} {\bibfield
  {journal} {\bibinfo  {journal} {Phys. Rev. E}\ }\textbf {\bibinfo {volume}
  {90}},\ \bibinfo {pages} {022302} (\bibinfo {year} {2014})}\BibitemShut
  {NoStop}%
\bibitem [{\citenamefont {Sentjabrskaja}\ \emph {et~al.}(2013)\citenamefont
  {Sentjabrskaja}, \citenamefont {Babaliari}, \citenamefont {Hendricks},
  \citenamefont {Laurati}, \citenamefont {Petekidis},\ and\ \citenamefont
  {Egelhaaf}}]{sentjabrskaja2013}%
  \BibitemOpen
  \bibfield  {author} {\bibinfo {author} {\bibfnamefont {T.}~\bibnamefont
  {Sentjabrskaja}}, \bibinfo {author} {\bibfnamefont {E.}~\bibnamefont
  {Babaliari}}, \bibinfo {author} {\bibfnamefont {J.}~\bibnamefont
  {Hendricks}}, \bibinfo {author} {\bibfnamefont {M.}~\bibnamefont {Laurati}},
  \bibinfo {author} {\bibfnamefont {G.}~\bibnamefont {Petekidis}}, \ and\
  \bibinfo {author} {\bibfnamefont {S.}~\bibnamefont {Egelhaaf}},\ }\href@noop
  {} {\bibfield  {journal} {\bibinfo  {journal} {Soft Matter}\ }\textbf
  {\bibinfo {volume} {9}},\ \bibinfo {pages} {4524} (\bibinfo {year}
  {2013})}\BibitemShut {NoStop}%
\bibitem [{\citenamefont {Jin}\ and\ \citenamefont {Yoshino}(2017)}]{Jin2017}%
  \BibitemOpen
  \bibfield  {author} {\bibinfo {author} {\bibfnamefont {Y.}~\bibnamefont
  {Jin}}\ and\ \bibinfo {author} {\bibfnamefont {H.}~\bibnamefont {Yoshino}},\
  }\href {http://dx.doi.org/10.1038/ncomms14935} {\bibfield  {journal}
  {\bibinfo  {journal} {Nature Communications}\ }\textbf {\bibinfo {volume}
  {8}},\ \bibinfo {pages} {14935} (\bibinfo {year} {2017})}\BibitemShut
  {NoStop}%
\bibitem [{Note3()}]{Note3}%
  \BibitemOpen
  \bibinfo {note} {In the mean-field p-spin spherical model, the MCT is
  unambiguously the dynamical counterpart of the replica theory~\cite
  {castellani2005spin}. In the infinite dimensional monodisperse particles, the
  dynamical transition predicted by the replica theory share the common
  features with the one predicted by the MCT~\cite
  {parisi2010,Maimbourg2016}.}\BibitemShut {Stop}%
\bibitem [{\citenamefont {Castellani}\ and\ \citenamefont
  {Cavagna}(2005)}]{castellani2005spin}%
  \BibitemOpen
  \bibfield  {author} {\bibinfo {author} {\bibfnamefont {T.}~\bibnamefont
  {Castellani}}\ and\ \bibinfo {author} {\bibfnamefont {A.}~\bibnamefont
  {Cavagna}},\ }\href@noop {} {\bibfield  {journal} {\bibinfo  {journal}
  {Journal of Statistical Mechanics: Theory and Experiment}\ }\textbf {\bibinfo
  {volume} {2005}},\ \bibinfo {pages} {P05012} (\bibinfo {year}
  {2005})}\BibitemShut {NoStop}%
\end{thebibliography}%

\end{document}